\documentclass[reprint,amsmath,amssymb,aps,pre,fleqn,showpacs]{revtex4-1}
\usepackage{graphicx}
\usepackage{enumerate}
\usepackage{dcolumn}
\usepackage{bm}
\begin{document}
\title{Capacitively-coupled rf discharge with a large amount of microparticles: 
spatiotemporal emission pattern and microparticle arrangement}
\author{M.$\,$Y.$\,$Pustylnik}
\email{mikhail.pustylnik@dlr.de}
\author{I.$\,$L.$\,$Semenov}
\author{E.$\,$Z\"{a}hringer} 
\author{H.$\,$M.$\,$Thomas}
\affiliation{Institut f\"ur Materialphysik im Weltraum, Deutsches Zentrum f\"{u}r Luft- und Raumfahrt (DLR), 82234 We\ss ling, Germany}

\date{\today}
\begin{abstract}
The effect of micron-sized particles on a low-pressure capacitively-coupled rf discharge is studied both experimentally and using numerical simulations. In the laboratory experiments, microparticle clouds occupying a considerable fraction of the discharge volume are supported against gravity with the help of the thermophoretic force. The spatiotemporally resolved optical emission measurements are performed with different arrangements of microparticles. The numerical simulations are carried out on the basis of a one-dimensional hybrid (fluid-kinetic) discharge model describing the interaction between plasma and microparticles in a self-consistent way. The study is focused on the role of microparticle arrangement  in interpreting the spatiotemporal emission measurements. We show that
it is not possible to reproduce simultaneously the observed microparticle arrangement and emission pattern in the framework of the considered one-dimensional model. This disagreement is discussed and attributed to two-dimensional effects, e.g., radial diffusion of the plasma components.
\end{abstract}


\maketitle
\newcommand{\D}{\partial}
\newcommand{\io}{_{\rm i}}
\newcommand{\el}{_{\rm e}}
\newcommand{\ds}{_{\rm p}}
\newcommand{\gs}{_{\rm g}}
\newcommand{\id}{_{\rm ip}}
\newcommand{\ed}{_{\rm ep}}
\newcommand{\ig}{_{\rm ig}}
\section{Introduction}
Low-pressure capacitively coupled rf discharges containing micro- or nanometer-sized particles have attracted a lot of attention in the last decades.~This interest was mainly driven by technological applications of such systems, e.g., for nanoparticle growth in plasmas \cite{mikikian2007self,couedel2010self,van2015fast, wegner2016influence,van2016conclusive,pilch2017diagnostics}. 
It has been shown long ago that the presence of nanoparticles leads to dramatic changes in plasma properties \cite{boehm1991silanerf, bouchoule1993highconcentration, fridman1996dustgrowththeory}.\\ \indent
Cold low-pressure plasmas with micron-sized particles have also found their application in basic science. 
They are used to create the so-called complex plasmas \cite{morfill2009complex,bonitz2010complex}, where the subsystem of strongly-coupled microparticles is easily accessible for observations at the atomistic level. 
Various generic phenomena of condensed matter physics can be studied using complex plasmas. 
To achieve large three-dimensional microparticle clouds,  complex plasma experiments are often performed under microgravity conditions \cite{thomas2008complex, pustylnik2016pk4}, where the microparticles occupy a significant part of the discharge volume. 
In this case their interaction with the background plasma becomes  very important.
Development of future microgravity plasma experiments would require fighting back the present drawbacks, namely strong inhomogeneity of the microparticle clouds (due to the microparticle-free areas, so-called voids \cite{thomas2008complex}) and low-frequency instabilities of the microparticle component (e.g., heartbeat instability \cite{heidemann2011hb, pustylnik2012heterogeneous}). 
This challenging goal can hardly be achieved without fundamental understanding of the interaction between the microparticle clouds and plasmas.\\ \indent
In recent years, this interaction has been studied both experimentally and using numerical simulations in several works \cite{hubner2009dust,mitic2009spectroscopic,melzer2011phase,killer2013observation}. 
Experimentally, the deepest insight into the plasma kinetics can be achieved by rf-period-resolved emission measurements.~Being non-invasive, this method is very well suited for the investigations of the rf discharges with microparticles. 
Particle-in-cell (PIC) simulations are the most common technique that also allows to access plasma kinetics. 
Both rf-period-resolved measurements and PIC simulations were combined in Refs.~\cite{melzer2011phase,killer2013observation} for the conditions of complex plasmas.\\ \indent
In Ref.~\cite{killer2013observation}, large microparticle clouds occupying the most part of the discharge volume were considered.~It has been shown that the increase in the microparticle number density causes the transition in the electron heating mode. 
While the electrons in the microparticle-free discharge are primarily heated in the sheath regions, a nearly uniform distribution of the electron heating rate across the discharge is observed in the presence of dense microparticle clouds.
This effect, which was also reported for electronegative plasmas \cite{schulze2011ionization,schungel2013effect,hemke2012ionization}, was attributed to the reduction of plasma conductivity due to the depletion of electrons within a microparticle cloud. 
The numerical simulations of Ref.~\cite{killer2013observation} confirmed this mechanism for the complex plasma conditions and showed satisfactory agreement with the rf-period-resolved emission measurements.\\ \indent
On the other hand, the effect of microparticle arrangement on the plasma properties and observed emission pattern was not discussed in Ref.~\cite{killer2013observation}. 
In fact, the simulations of Ref.~\cite{killer2013observation} were performed for a prescribed fixed distribution of the microparticle number density. 
The presence of the void in the center of the discharge was not taken into account, whereas (as already mentioned above) lack of the fundamental understanding of the void constitutes a major obstacle in the improvement of microgravity plasma experiments.
\\ \indent
In the present work, we provide some initial insights into the role of the details of the microparticle arrangement in the physics of the discharge containing a large amount of microparticles. 
In particular, we report on the results of experiments similar to those performed in Ref.~\cite{killer2013observation}: measured space- and time-resolved emission patterns are compared with the results of numerical simulations. 
The simulations are performed on the basis of a one-dimensional hybrid (fluid-kinetic) model, in which the interaction of plasma with microparticles is taken into account self-consistently. 
This allows us to predict the microparticle cloud configuration. 
For comparison, we also demonstrate the results of simulations with the fixed microparticle arrangement. 
Finally, the findings of our study are used to discuss the physical mechanisms underlying the formation and existence of the void usually observed in the center of the discharge.\\ \indent
The paper is structured as follows. In Sec.~\ref{sec: Exp} we describe the experimental setup used in our work and the details of the rf-period-resolved emission measurements.  
The model used for numerical simulations is presented in Sec.~\ref{Sec: Model}. Our main findings are summarized in Sec.~\ref{sec: Results}. 
Discussion of the results is given in Sec.~\ref{Sec: Discuss} and the conclusions are drawn in Sec.~\ref{Sec: Concl}.
\section{Experiment}
\label{sec: Exp}
\begin{figure}
\centering
\includegraphics[width=0.35\textwidth]{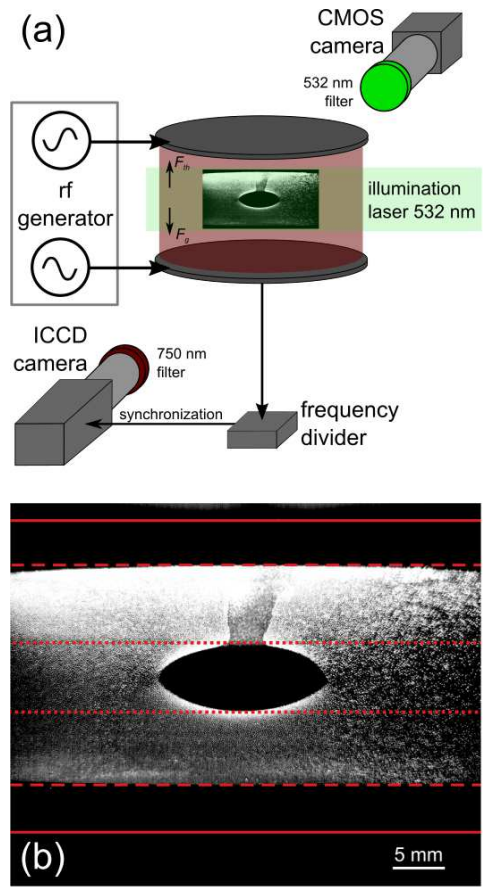}
\caption{\label{Fig: Scheme} 
(a) Sketch of the experimental setup. RF argon plasma is generated between the two electrodes powered in push-pull mode. 
The spatiotemporal profile of the plasma emission is observed by an ICCD camera, which is synchronized with the rf voltage via a frequency divider. 
Microparticles are injected into the plasma, illuminated by a laser sheet and observed by a CMOS camera. 
Large volumetric microparticle clouds are supported against gravity by the thermophoretic force. 
(b) Typical appearance of a microparticle cloud with a central void. 
Solid, dashed and dotted lines mark the positions of the electrodes, edges of the microparticle cloud and axial extension of the void, respectively.}
\end{figure}
\subsection{Plasma reactor}
A sketch of the experimental setup is presented in Fig.~\ref{Fig: Scheme}(a).
The experiments were conducted in the PK-3~Plus chamber \cite{thomas2008complex}, which is a compact parallel-plate plasma reactor (the interelectrode gap is $3\,$cm, the electrode diameter is $6\,$cm).
The plasma was produced by means of a capacitively-coupled rf discharge. 
Two electrodes were driven in push\mbox{-}pull mode by a sinusoidal signal with the frequency of $13.56\,$MHz and amplitude of about $100\,$V peak\mbox{-}to\mbox{-}peak.
Argon fed to the chamber with $0.2$~sccm gas flow was used as working gas.
Pressure was controlled in the range of $20$-$40\,$Pa. 
\subsection{Microparticles}
\label{Sec: MuPart}
Monodisperse plastic microspheres of $1.95\,\mu$m diameter were levitated in the plasma. 
Under ground laboratory conditions the microparticles concentrate themselves in the vicinity of the lower electrode. 
To obtain large volumetric microparticle clouds, we compensated the gravitational force by means of themophoresis \cite{rothermel2002thermophoresis}. 
To achieve this, we controlled the temperature difference between the bottom and top flanges \cite{schwabe2007highly} keeping the bottom flange about $17\,$K hotter than the top one.
The microparticle cloud in the plasma was visualized by illuminating it with a sheet of a green (wavelength $532\,$nm) laser entering the chamber in its middle. 
Scattered laser light (filtered out by a respective interference filter) was detected by a CMOS videocamera with the resolution of $30\,\mu$m/pix. 
\\ \indent 
Typical appearance of a microparticle cloud in our experiments is shown in Fig.~\ref{Fig: Scheme}(b). 
As under microgravity conditions \cite{thomas2008complex}, the cloud has a characteristic void in its center. 
Dashed and dotted lines show the edges of the cloud and the axial void extension, respectively.
Below we will refer to these two features for the explanation and discussion of the experimental results.
\subsection{Emission measurements}
\label{Sec: ExpEmission}
\begin{figure}[!h]
\includegraphics[width=0.4\textwidth]{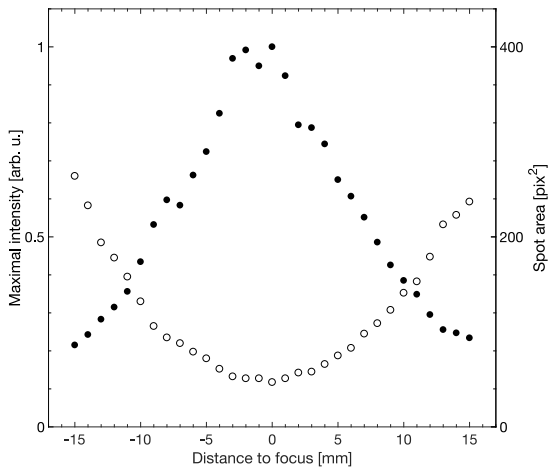}
\caption{\label{Fig: Defoc} 
Depth of field investigation of the ICCD camera: dependence of the spot area (open circles) and maximal pixel intensity (full circles) in the image of a point-like source on the distance between the point-like source and camera focus. 
Negative abscissa displacement corresponds to the displacement towards the ICCD camera.}
\end{figure}

RF-period-resolved evolution of the plasma emission was observed by an ICCD camera.
Following previous studies \cite{melzer2011phase,killer2013observation}, the emission pattern was analyzed for the argon lines 750.4$\,$nm and 751.5$\,$nm. 
These lines correspond to the transitions $2\rm{p}_1\to1\rm{s}_2$ and $2\rm{p}_5\to1\rm{s}_4$ (in Paschen notation) with lifetimes  22.5$\,$ns and 24.9$\,$ns, respectively \cite{wiese1989unified}. 
To select these lines, a filter with $750\,$nm central wavelength and $10\,$nm width was placed in front of the camera lens.
The camera was synchronized with the rf signal through the frequency divider which sent a synchronization pulse after every $512$ rf cycles. 
The gate width of the ICCD camera was set to $10\,$ns. The signal was accumulated over the total exposure time of $1\,$s. 
The ICCD gate was moved over almost three rf periods with a step of $2\,$ns. 
The spatial resolution of the ICCD camera was $75\,\mu$m/pix.
\\ \indent
Since the plasma is extended along the line-of-sight of the ICCD camera, the observed emission pattern is affected, to some extent, by the line-of-sight averaging. 
In order to estimate the influence of this averaging, the depth of field of the ICCD camera was characterized. 
This was done by imaging the same point-like light source placed at different distances from the camera lens. In these tests, $100\,$mm camera lens and $750\,$nm filter were used. 
The same optical system (camera, lens and filter) was used later in all experiments. 
The output of a $25\,\mu$m diameter optic fiber was used as a point-like source. 
The optic fiber was connected to a tungsten-halogen lamp on its other end. 
First, this source was placed at a distance of $578\,$mm from the camera lens. The lens focus was adjusted to minimize the spot size in the image. 
Then, the source was moved  $\pm15\,$mm from the initial position with a step of $1\,$mm using a translation stage. 
At every step, an image was taken. 
The results of this test are summarized in Fig.~\ref{Fig: Defoc} in terms of two image parameters: the spot area and maximal pixel intensity. 
This information allows to judge about the contribution of different plasma layers into the line-of-sight averaged picture.\\ \indent
For the experiments, the ICCD camera was placed at the same distance of $578\pm2\,$mm from the illumination laser sheet plane. 
The same focusing of the camera lens as in the depth-of-field test was kept.
\section{Model}
\label{Sec: Model}
\subsection{Basic equations}
Numerical simulations were performed using a one-dimensional (parallel-plate) model of the discharge.
The plasma was assumed to consist of argon atoms, electrons, singly charged argon ions and charged microparticles.
The atoms were treated as a uniform background at temperature $T\gs=300\,$K. 
The electron component was described using the conventional particle-in-cell Monte-Carlo (PIC-MCC) simulation technique \cite{vahedi1995monte,donko2011particle}. 
The ion component was described by means of the fluid model based on the moment equations proposed in Ref.~\cite{semenov2016moment}. 
The microparticles were treated as a fluid within the drift-diffusion model considered previously in a number of works \cite{gozadinos2003fluid,land2006effect,land2007plasma,goedheer2009hydrodynamic,land2010fluid}.
The details of the model for each component are given below.
\\ \indent
The adopted Monte-Carlo model of the electron-atom and electron-microparticle collisions was based on that of Ref.~\cite{semenov2016moment}. 
The absorption of the electrons by the microparticles was modeled as an additional collision process with the orbit-motion-limited (OML) \cite{fortov2005complex} cross-section. 
In addition, the electron scattering model was improved. Namely, the differential cross-section for the electron-atom collisions was calculated using the expression known from the quantum scattering theory~\cite{landau1965course}. 
The corresponding phase shifts were taken from Ref.~\cite{mceachran1983elastic}.
\\ \indent
The governing equations for ions included the continuity, momentum and energy equations \cite{semenov2016moment}.
The continuity equation reads
\begin{align}
\label{IonCont}
\frac{\D n\io}{\D t}+\frac{\D }{\D x}(n\io u\io)=G_{\rm ion}-G_{\rm ip},
\end{align}
where $t$ is time, $x$ is the coordinate along the axis perpendicular to the electrodes, $n\io$ is the ion number density, $u\io$ is the ion mean velocity, $G_{\rm ion}$ is the ionization source and $G\id$ is the sink of ions due to absorption by the microparticles.
The ion momentum equation reads
\begin{align}
\label{IonMoment}
\frac{ \D (m\io n\io u\io) }{\D t}+\frac{\D}{\D x}
\left( m\io n\io u\io^{2}+p\io \right)=en\io E \nonumber \\ 
-\omega\ig m\io n\io u\io-n\io n\ds f\id,
\end{align}
where $m\io$ is the ion mass, $p\io$ is the ion pressure,  $e$ is the absolute value of the elementary charge, $E$ is the electric field,  $\omega \ig$ is the momentum-transfer frequency for the ion-atom collisions, $n\ds$ is the microparticle number density and $f\id$ is the ion drag force normalized to the ion number density.
The evolution of the ion pressure was governed by the energy equation
\begin{align}
\label{IonEenergy}
\frac{\D \Pi \io}{\D t}+
\frac{\D}{\D x} \left( m\io n\io u\io^{3} +3 \, p\io u\io  \right)= 2n\io u\io eE \nonumber \\
+\omega \ig \delta \left( kT\gs -\Pi\io \right),
\end{align}
where $\Pi\io= p\io +m\io n\io u\io^{2}$ and $\delta$ is a correction coefficient introduced in Ref.~\cite{semenov2016moment}. 
Note that the effect of the ion-microparticle collisions is neglected in Eq.~(\ref{IonEenergy}) due to the large difference between the ion mass and the mass of the microparticle.
The definition of the effective collision frequency $\omega \ig$ can be found in Ref.~\cite{semenov2016moment}.\\ \indent
The terms $G_{\rm ion}$, $G_{\rm ip}$ and the ion drag force were calculated using the conventional expressions presented in Ref.~\cite{fortov2005complex}. 
The collection cross-section for the ion-microparticle collisions was defined using the OML theory. 
The effect of the ion-atom collisions on the ion absorption rate was taken into account using the model of Ref.~\cite{khrapak2008interpolation}. 
The momentum transfer cross-section for ion scattering on the microparticles was evaluated using the expression from Ref.~\cite{khrapak2014accurate}. 
The nonlinear screening effects were taken into account by introducing the effective screening length \cite{ratynskaia2006electrostatic,semenov2015approximate,
semenov2017momentum}. 
Following the results of Ref.~\cite{semenov2016moment}, the ion velocity distribution function was defined using the model of Ref.~\cite{lampe2012ion} (applied for the local mean velocity $u\io$).
\\ \indent
The evolution of the microparticle number density was modelled by the following drift-diffusion equation:
\begin{align}
\label{DustDD}
\frac{\D n\ds}{\D t}+
\frac{\D}{\D x}(u\ds  n\ds)=
\frac{\D^{2}}{\D x^{2}} \left( D\ds n\ds \right),
\end{align}
where $D_{\ds}$ is the microparticle diffusion coefficient and
\begin{align}
\label{Vd}
u \ds=(q \ds   E+n\io f\id)/ \omega \ds m\ds,
\end{align}
is the microparticle drift velocity with $q \ds$ being the microparticle charge, 
$m \ds$ being the microparticle mass and $\omega \ds$ being the momentum transfer frequency for the microparticle-atom collisions. 
The latter was evaluated using the conventional Epstein relation \cite{fortov2005complex}. 
The gravitational and thermophoretic forces on the microparticle are assumed to compensate each other and, therefore, are ignored in Eq.~(\ref{Vd}).
The diffusion coefficient is given by
\begin{align}
\label{Dd}
D\ds=k T\ds / \omega\ds m\ds,
\end{align}
where $k$ is the Boltzmann constant and $T\ds$ is the effective microparticle temperature which can be calculated using an appropriate thermodynamic model \cite{khrapak2014ion,khrapak2014simple,khrapak2015practical}.
In the present work $T\ds$ was evaluated within the model proposed in Ref.~\cite{khrapak2015practical}. 
Following the results of Ref.~\cite{khrapak2015practical}, the expression for the microparticle temperature is given by
\begin{align}
\label{Td}
T\ds=T_{\rm g} \left( 1+ \frac{\kappa^{4} \Gamma }{6 \left [ 
\kappa \cosh(\kappa)-\sinh(\kappa)
\right ]^{3}} \right),
\end{align}
where $\kappa$ and $\Gamma$ are the screening and coupling parameters, respectively. 
The screening and coupling parameters are defined as $\kappa=a/ \lambda_{\rm D}$ and $\Gamma=q\ds^{2}/akT\gs$ , where $a=(3/4 \pi n\ds)^{1/3}$ is the Wigner-Seitz radius and $\lambda_{\rm D}$ is the Debye length in plasma. 
The second term in Eq.~(\ref{Td}) accounts for the interaction between the microparticles. 
This term was derived in Ref.~\cite{khrapak2014ion} using the ion-sphere model for Yukawa systems and was shown to give an accurate estimate of the effective microparticle temperature both for crystalline and fluid phases. 
The temperature of noninteracting particles, as it follows from Eq.~(\ref{Td}), equals to that of the background gas.\\ \indent
The microparticle charge was assumed to depend on the local plasma parameters. The local values of the particle charge were calculated using the charging equation
\begin{align}
\label{Qd}
n\ds \, \frac{\D q\ds}{\D t}=e (G\id-G\ed),
\end{align}
where $G_{\rm ep}$ is the electron flux to the microparticle \cite{fortov2005complex}. \\ \indent
The model of the discharge was closed by the Poisson equation for the self-consistent electric field:
\begin{align}
\label{Poisson}
\Delta \varphi=-4 \pi (en\io-en\el-q\ds n\ds),
\end{align}
where $\varphi$ is the electric field potential and $n\el$ is the electron number density. 
The value of $\varphi$ was set to zero at $x=L$ ($L$ is the distance between the electrodes) and was varied sinusoidally (frequency 13.56$\,$MHz) in time at $x=0$.
\subsection{Numerical procedure}
The governing equations (\ref{IonCont})-(\ref{DustDD}) were solved numerically by means of the conventional finite-volume methods developed in the computational fluid dynamics \cite{leveque2002finite}. 
The electrodes were assumed to absorb all incident ions and electrons. Eq.~(\ref{DustDD}) was solved with the zero-flux boundary conditions. 
The boundary conditions for Eqs.~(\ref{IonCont})-(\ref{IonEenergy}) were defined as described in Ref.~\cite{semenov2016moment}. 
The Poisson equation~(\ref{Poisson}) was approximated using the second-order finite-difference method. The number of mesh points between the electrodes was set to 400. 
The final number of particles used in the PIC-MCC model for electrons was of the order of $10^{5}$. 
The number of time steps within the rf period was set to 4000.
\\ \indent
In all simulations, the initial plasma state was defined as follows. 
The ions and electrons were assumed to have non-shifted Maxwellian velocity distributions. 
The initial electron temperature and initial plasma density were estimated using a global model of the pure discharge (without microparticles). 
The initial ion temperature was set to $T_{\rm g}$. 
The initial distribution of the microparticle number density was chosen to mimic the distribution of the microparticles on the discharge axis (see Fig.~\ref{Fig: Scheme}(b)). 
The microparticle number density was set to zero in the regions near the electrodes and inside the void.
In the two remaining intervals, the microparticle number density was assumed to have a uniform distribution (blue line in Fig.~\ref{Fig: DustDensity}).
The initial values of the local microparticle charge were set to zero.
The simulations were run until periodic steady-state was reached.
\\ \indent
The main difficulty inherent to the considered problem is the large difference between the time scales relevant to the evolution of the microparticle cloud and other plasma components.
In view of this difficulty, realistic simulations of the microparticle dynamics represent a challenging computational problem. 
On the other hand, the steady-state solutions of Eq.~(\ref{DustDD}) does not depend on the parameter $\omega \ds m \ds$. 
Therefore, if only a steady-state microparticle configuration is of interest, the simulations can be performed with some effective (reduced) value of $\omega \ds m\ds$ \cite{gozadinos2003fluid}. 
According to this approach, in our work, the simulations were performed with the parameter $\omega \ds m \ds$ reduced by a factor of $10^{4}$. 
Several test simulations with different values of the reducing factor ($10^{3}$ and $10^{5}$) showed that the final steady-state distribution of $n\ds$ is practically independent of the chosen value of $\omega \ds m \ds$.
\\ \indent
It should be noted that the characteristic time required to reach the steady-state microparticle number density profile in the simulations with the reduced value of 
$\omega \ds m \ds$ was of the order of several thousands of rf periods.
This time is much longer than the characteristic time scale of the transport processes in the pure plasma. 
Thus, a characteristic separation of the time scales in the considered system was qualitatively reproduced in our simulations.
In addition, our numerical code was tested to reproduce the void closure experiments performed with a relatively small number of microparticles \cite{lipaev2007void}. 
In that case, the effect of microparticles on the plasma properties was negligible.
\\ \indent
Using the obtained simulation results, the period resolved emission pattern for the selected argon lines (see Sec.~\ref{sec: Exp}) was compared with the experimental data. 
The emission pattern was calculated by the convolution of the excitation rate and the decay function $\exp(-t/\tau)$, where $\tau$ is the lifetime of the transition. 
The excitation rates were calculated using the electron distribution function obtained from the simulations. 
The corresponding excitation cross-sections were taken from the Biagi-v7.1 database \citep{LXCat}.
\section{Results}
\label{sec: Results}
\subsection{Variation of the microparticle amount}
\label{Sec: results_variation}
\begin{figure}[!h]
\includegraphics[width=0.35\textwidth]{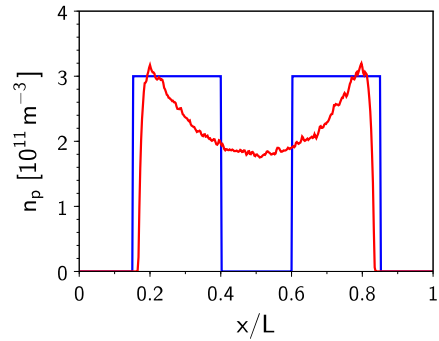}
\caption{\label{Fig: DustDensity} Typical spatial profiles of the microparticle number density: initial input profile (blue line) and self-consistent steady-state profile (red line).}
\end{figure}

\begin{figure*}
\includegraphics[width=0.9\textwidth]{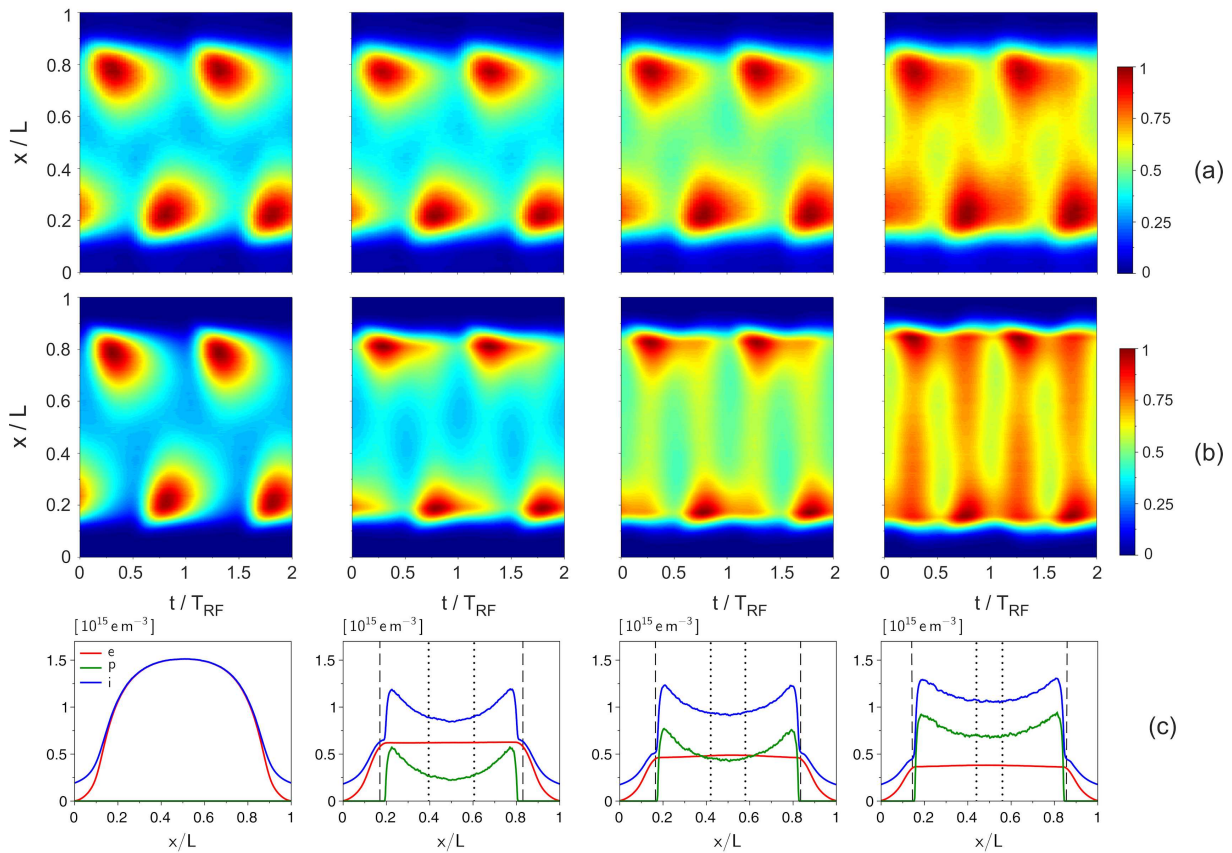}
\caption{\label{Fig: Iems_exp_num} 
Spatiotemporal emission patterns of an rf discharge in argon at $27\,$Pa and $ 54\,$V rf amplitude with increasing (from left to right) microparticle number density: (a) experiment and (b) simulations.
The leftmost  figures correspond to the case of pure plasma. 
The emission intensity is given in arbitrary units. (c) The rf\mbox{-}period\mbox{-}averaged charge density distributions for the electrons (red line), microparticles (green line) and ions (blue line) obtained from the simulations. 
The results correspond to the emission patterns shown in Fig.~\ref{Fig: Iems_exp_num}(b). 
The dashed lines show the boundaries of the microparticle cloud and the dotted lines show the axial void extension observed in the experiments (see Fig.~\ref{Fig: Scheme}(b)).}
\end{figure*}

Let us first discuss how the rf-period-resolved emission pattern is influenced by changes in the number density of microparticles. 
In the simulations, the initial level of the microparticle number density (plateau of the blue line in Fig.~\ref{Fig: DustDensity}) was varied in the range between $10^{11}$ and 
$5 \times 10^{11}\,\mathrm{m}^{-3}$. 
Taking into account that a typical microparticle charge for the present problem is of the order of $3000e$, one can see that the microparticle charge density for the chosen range of $n\ds$ is comparable to the charge density in the microparticle-free discharge (see Fig.~\ref{Fig: Iems_exp_num}(c)). 
In the experiments, the microparticle number density was successively increased by additional injection of microparticles.\\ \indent
The experimental measurements showed that the gas pressure variations do not affect qualitative  evolution of the emission pattern with increasing number of  microparticles.
For this reason, we focused on the results obtained at the gas pressure of 27$\,$Pa. 
In this case, the average voltage amplitude measured in the experiments was 54$\,$V.\\ \indent
In Figs.~\ref{Fig: Iems_exp_num}(a)-(b) we present some selected emission patterns observed in the experiments and obtained from the simulations with different number densities of the  microparticles. 
As it can be seen from Figs.~\ref{Fig: Iems_exp_num}(a)-(b), the model employed in the present work is able to reproduce, at least qualitatively, the experimental observations: the increase in the microparticle number density leads to a more uniform spatial distribution of the emission intensity.
This observation agrees with the findings of Ref.~\cite{killer2013observation} and is known to be caused by the growth of the electric field oscillation amplitude in the plasma bulk. 
The latter effect is connected with the decrease in plasma conductivity, which is a consequence of the electron depletion by the microparticles.  
The same phenomenon was previously observed in electronegative discharges \cite{schulze2011ionization,schungel2013effect,hemke2012ionization}, where the negative ions immobilize the negative charge in a similar way as the microparticles do it in complex plasmas.
\\ \indent
The above discussion can be supported by considering the charge density distributions for different plasma components. 
The corresponding numerical results for the selected emission patterns are shown in Fig.~\ref{Fig: Iems_exp_num}(c). 
As it can be seen, the emission pattern becomes more uniform as the microparticle charge density in the bulk region increases and the electron charge density decreases. 
On the other hand, as it follows from Fig.~\ref{Fig: Iems_exp_num}(c), the ion charge density in the sheath regions does not change significantly. 
Therefore, the displacement current induced in the oscillating rf sheath remains virtually constant with increasing microparticle number density.
As a result, the total current in the system is not strongly affected by the microparticles. 
This fact, in combination with the reduction of plasma conductivity due to the electron depletion, explains the increase in the electric field oscillation amplitude and respective growth of the emission intensity in the bulk.
\\ \indent
The distributions of the microparticle charge density presented in Fig.~\ref{Fig: Iems_exp_num}(c) are consistent with those observed in previous works devoted to the numerical studies of the microparticle arrangement \cite{sukhinin2013influence,sukhinin2013dust,land2006effect,
land2007plasma,goedheer2009hydrodynamic,land2010fluid}. 
In addition, we show in Fig.~\ref{Fig: Iems_exp_num}(c) the boundaries of the regions occupied by the microparticles in the experimental images near the discharge central axis. 
The comparison with the experimental data shows that our model reproduces well the positions of the microparticle cloud boundaries. 
The main disagreement with the experimental observations is connected with the existence of the central void. 
While the void is visible in all experimental images, it is not present in the microparticle number density distributions obtained from the one-dimensional simulations.
\subsection{Plasma emission inside the void}
\label{Sec: VoidEms}
\begin{figure}
\includegraphics[width=0.4\textwidth]{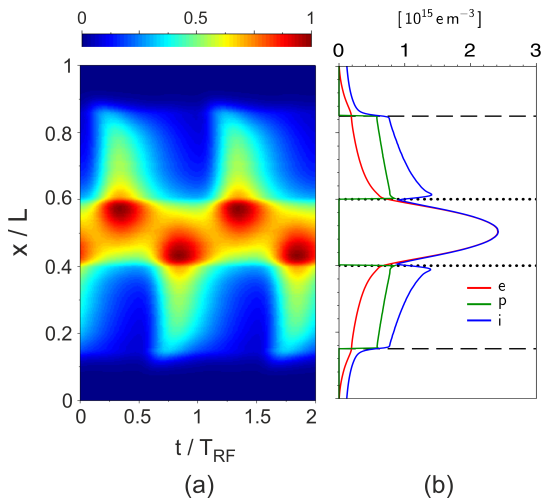}
\caption{\label{Fig: Iems_void} 
The results of the simulation run performed for a fixed microparticle number density distribution with the void in the central region (blue line in Fig.~\ref{Fig:  DustDensity}). 
The discharge parameters are the same as in Fig.~\ref{Fig: Iems_exp_num}. 
(a)~Spatiotemporal emission pattern. 
The emission intensity is given in arbitrary units. 
(b)~The rf\mbox{-}period\mbox{-}averaged charge density distributions for the electrons (red line), microparticles (green line) and ions (blue line). 
The dashed lines show the boundaries of the microparticle cloud and the dotted lines show the void extension defined in the simulation (see Fig.~\ref{Fig: DustDensity}).}
\end{figure}
To elucidate the possible effect of the void on the emission pattern we performed an additional simulation run with a fixed microparticle number density distribution which mimics that observed in the experiments (blue line in Fig.~\ref{Fig: DustDensity}).
~The equation~(\ref{DustDD}) was not considered in this case.
\\ \indent
The obtained results are summarized in Fig.~\ref{Fig: Iems_void}.~As it can be seen, the emission pattern changes drastically with respect to those in Fig.~\ref{Fig: Iems_exp_num}(b). 
The emission coming from the void noticeably dominates over the emission coming from the other parts of the discharge. This contradicts the experimental observations, where no visible effect of the void on the emission pattern is detected (see Fig.~\ref{Fig: Iems_exp_num}(a)).\\ \indent
On the other hand, the emission pattern presented in Fig.~\ref{Fig: Iems_void}(a) correlates with the calculated distribution of the electron number density.
In fact, the increase of the emission intensity in the void is connected with the increase of the electron number density in this region (see Fig.~\ref{Fig: Iems_void}(b)). The latter is simply a consequence of plasma quasineutrality and results from the abrupt reduction of the microparticle number density.
\subsection{Effect of the void position}
\begin{figure}
\includegraphics[width=0.49\textwidth]{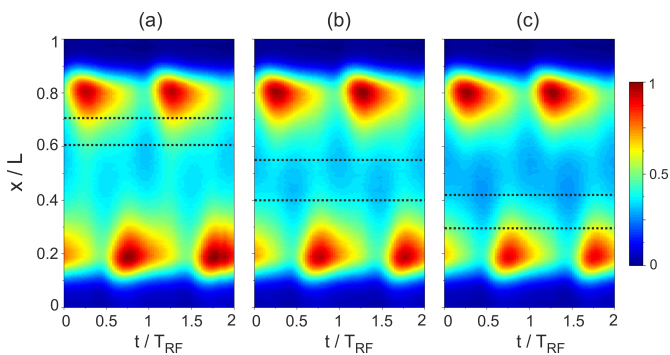}
\caption{\label{Fig: Iems_void_position} 
Spatiotemporal emission patterns observed in the experiments with different positions of the void. 
The dotted lines show the experimentally observed axial void extension (Fig.~\ref{Fig: Scheme}(b)). 
Void position was varied by changing the temperature difference between the chamber flanges: (a) $12\,$K, (b) $13\,$K, (c) $15\,$K.}
\end{figure}
In addition, a row of experiments was conducted in an attempt to find the dependence of the emission pattern on the position of the void inside the microparticle cloud.
The experiments were performed at the gas pressure of 36$\,$Pa and the voltage amplitude of 53$\,$V.
The position of the void was varied by changing the temperature difference between the electrodes. Representative examples of the observed emission patterns are shown in Fig.~\ref{Fig: Iems_void_position}.
As it can be seen, the variation of the void position does not produce any significant effect on the emission pattern. 
\section{Discussion}
\label{Sec: Discuss}
\begin{figure}
\includegraphics[width=0.35\textwidth]{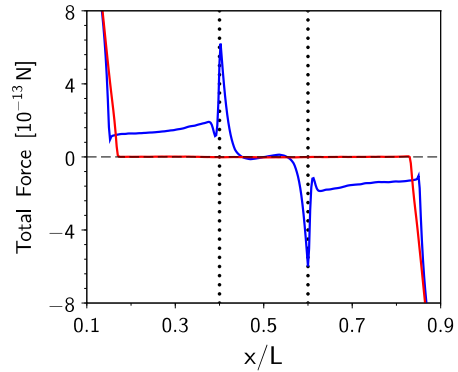}
\caption{\label{Fig: Force} 
The distribution of the total force acting on a microparticle obtained from two simulation runs. 
The blue line shows the force obtained from the simulation with a fixed microparticle number density distribution considered in Sec.~\ref{Sec: VoidEms}. 
The red line shows the force obtained from the simulation performed using a self-consistent model of the discharge. 
The dotted lines show the void boundaries defined for the simulation with a fixed microparticle number density distribution.}
\end{figure}
The results presented in Sec.~\ref{sec: Results} give rise to the following questions:
\begin{enumerate}[(i)]
\item Why is the void formation not reproduced by the numerical model? 
\item Why is the presence of the void not reflected in the experimentally observed spatiotemporal emission pattern?
\end{enumerate}
\subsection{Void closure}
In order to discuss the first question, we analyze the behavior of the total force acting on a microparticle in the discharge. 
In Fig.~\ref{Fig: Force} we show the distributions of this force obtained from two different simulation runs. 
In the first run, the microparticle number density distribution (blue line in Fig.~\ref{Fig: DustDensity}) was fixed.
In the second run, the same initial distribution was allowed to evolve self-consistently. 
The resulting microparticle number density distribution is given by the red line in Fig.~\ref{Fig: DustDensity}.
\\ \indent
As it can be seen from Fig.~\ref{Fig: Force}, the total force on a  microparticle exhibits a sharp peak in the vicinity of the void boundary. 
The electric field responsible for this peak appears as a consequence of the double-layer formation at the boundary between the plasmas with different electron thermal fluxes (see, e.g., Ref.~\cite{ishiguro1985dblayer}). 
This effect may be enhanced by the presence of the rf electric field perpendicular to the boundary \cite{raizer1995rfdischarge}. 
However, it is difficult to separate the two mechanisms
in the frame of a one-dimensional model.
\\ \indent
The discussion above indicates that the void closure seems to be an inevitable phenomenon for the considered one-dimensional problem. 
In fact, as long as the void exists, the double layer electric field at the void boundary will push the microparticles into the central region of the discharge. 
The microparticles can reach the equilibrium position near the void edge only if the electrostatic force exerted on a microparticle at the void boundary will be comparable to the ion drag force. 
This condition can be satisfied only when the effect of microparticles on the plasma is very small.
\\ \indent
In addition, it should be noted that the mechanism of the void closure described above differs from that proposed in Ref.~\cite{pustylnik2012heterogeneous} to explain the heartbeat instability observed under certain conditions. 
The double-layer in our case occurs as a consequence of the contact between two plasmas with different electron thermal fluxes. 
This phenomenon does not have a critical character: the double layer forms even if the plasmas in contact are only slightly different (i.e., if the microparticle number density is very small). 
In contrast to that, collapse of the void during the heartbeat instability is a critical phenomenon \cite{pustylnik2012heterogeneous} which occurs due to an abrupt change in the double layer structure caused by steadily increasing plasma absorption at the void boundary.
\\ \indent
As it can be also seen in Fig.~\ref{Fig: Force}, the total force on a microparticle calculated within the self-consistent model  is nearly zero inside the microparticle cloud and does not reveal any noticeable potential barriers for the microparticles in the central region of the discharge. 
It is worth noting that the same finding has been already reported in Ref.~\cite{sukhinin2013dust}.\\ \indent
To sum up, our results demonstrate that the existence of the void in our experiments can not be consistently explained by a one-dimensional model of the discharge.
This conclusion indicates that two-dimensional effects might play a key role in the formation of the microparticle clouds with a large amount of microparticles.
\subsection{Emission pattern}
Let us now discuss the second question stated at the beginning of this section. 
The results shown in Fig.~\ref{Fig: Iems_void} demonstrate that the existence of the fixed void within the considered one-dimensional model would provide a dramatic effect on the emission pattern.
A possible reason why it is not observed in the experimental data is the line-of-sight averaging. 
In fact, the observed emission pattern results from the contribution of different plasma layers containing the regions both with and without the void.
Under typical conditions, the maximal radial width of the void is approximately one third of the microparticle cloud diameter. 
Thus, the line-of-sight averaging effect cannot be neglected in this case. 
However, taking into account the depth-of-field characteristics of our optical system (see Fig.~\ref{Fig: Defoc}), it seems unlikely that the line-of-sight averaging  is able to completely suppress the characteristic emission spot predicted in Fig.~\ref{Fig: Iems_void}. 
Moreover, the emission coming from the void has been indeed detected under certain conditions \cite{samsonov1999instabilities,berndt2005anomalous,mikikian2007self,couedel2010self,pustylnik2012heterogeneous,van2015fast}. 
This suggests that in our conditions, a real three-dimensional void produces a much weaker disturbance of the spatiotemporal emission pattern than that shown in Fig.~\ref{Fig: Iems_void}.
\subsection{Two-dimensional effects}
As it was mentioned above, the two-dimensional effects might be crucial in understanding the discrepancies between the experimental and simulation results observed in our study. 
One important feature, which is not considered in the one-dimensional model (but is present in the experiment), is the radial diffusion of the plasma. 
The radial diffusion serving as an additional mechanism of void plasma losses could lead to the reduction of the electron number density inside the void. 
This would, in turn, lead to the reduction of the double layer electric field at the void boundary. 
Thus, it seems possible that under certain conditions the electrostatic force pushing the microparticles towards the void center could be compensated by the drag force exerted by outflowing ions. 
The reduction of the electron number density in the void could, in its turn, lead to the reduction of the influence of the void on the rf-period-resolved emission pattern to such extent, that the void plasma emission could be obscured by the line-of-sight averaging.
\\ \indent
\section{Conclusions}
\label{Sec: Concl}
In conclusion, the effect of micron-sized particles on a capacitively coupled rf discharge has been studied both experimentally and via numerical simulations. 
In the experiments, microparticle clouds were confined in an argon discharge with the help of thermophoretic levitation.~The space- and time-resolved emission measurements were performed with different amounts of microparticles.~The corresponding numerical simulations were carried out on the basis of a one-dimensional discharge model describing the plasma and microparticle components in a self-consistent way.\\ \indent
The main findings of our work can be summarized as follows. 
The spatiotemporal emission patterns observed in the experiments are qualitatively well reproduced by the self-consistent numerical simulations. 
However, there is a strong disagreement between the experimental and numerical results regarding the formation of the central void (microparticle-free region) and the influence of the void on the emission profile. 
In particular, the void formation is not reproduced by the self-consistent simulations. Moreover, it is shown that the void closure seems to be an inevitable effect in the framework of the considered one-dimensional model. In addition, the simulations with a fixed microparticle arrangement, which mimics the presence of the void, indicate that the void induced emission would noticeably affect the observed emission pattern.  
This, however, contradicts the experimental data.\\ \indent
The observed discrepancies between experimental and numerical results have been discussed and attributed to the two-dimensional effects (e.g., radial plasma diffusion), which are not taken into account by our model. 
This hypothesis can be verified by a more detailed investigation based on two-dimensional numerical simulations and active depth-resolved optical plasma diagnostics, e.g., laser-induced fluorescence.
\section{Acknowledgement}
The authors appreciate helpful discussions with A. Zobnin.

\begin{thebibliography}{54}%
\makeatletter
\providecommand \@ifxundefined [1]{%
 \@ifx{#1\undefined}
}%
\providecommand \@ifnum [1]{%
 \ifnum #1\expandafter \@firstoftwo
 \else \expandafter \@secondoftwo
 \fi
}%
\providecommand \@ifx [1]{%
 \ifx #1\expandafter \@firstoftwo
 \else \expandafter \@secondoftwo
 \fi
}%
\providecommand \natexlab [1]{#1}%
\providecommand \enquote  [1]{``#1''}%
\providecommand \bibnamefont  [1]{#1}%
\providecommand \bibfnamefont [1]{#1}%
\providecommand \citenamefont [1]{#1}%
\providecommand \href@noop [0]{\@secondoftwo}%
\providecommand \href [0]{\begingroup \@sanitize@url \@href}%
\providecommand \@href[1]{\@@startlink{#1}\@@href}%
\providecommand \@@href[1]{\endgroup#1\@@endlink}%
\providecommand \@sanitize@url [0]{\catcode `\\12\catcode `\$12\catcode
  `\&12\catcode `\#12\catcode `\^12\catcode `\_12\catcode `\%12\relax}%
\providecommand \@@startlink[1]{}%
\providecommand \@@endlink[0]{}%
\providecommand \url  [0]{\begingroup\@sanitize@url \@url }%
\providecommand \@url [1]{\endgroup\@href {#1}{\urlprefix }}%
\providecommand \urlprefix  [0]{URL }%
\providecommand \Eprint [0]{\href }%
\providecommand \doibase [0]{http://dx.doi.org/}%
\providecommand \selectlanguage [0]{\@gobble}%
\providecommand \bibinfo  [0]{\@secondoftwo}%
\providecommand \bibfield  [0]{\@secondoftwo}%
\providecommand \translation [1]{[#1]}%
\providecommand \BibitemOpen [0]{}%
\providecommand \bibitemStop [0]{}%
\providecommand \bibitemNoStop [0]{.\EOS\space}%
\providecommand \EOS [0]{\spacefactor3000\relax}%
\providecommand \BibitemShut  [1]{\csname bibitem#1\endcsname}%
\let\auto@bib@innerbib\@empty
\bibitem [{\citenamefont {Mikikian}\ \emph {et~al.}(2007)\citenamefont
  {Mikikian}, \citenamefont {Cou{\"e}del}, \citenamefont {Cavarroc},
  \citenamefont {Tessier},\ and\ \citenamefont {Boufendi}}]{mikikian2007self}%
  \BibitemOpen
  \bibfield  {author} {\bibinfo {author} {\bibfnamefont {M.}~\bibnamefont
  {Mikikian}}, \bibinfo {author} {\bibfnamefont {L.}~\bibnamefont
  {Cou{\"e}del}}, \bibinfo {author} {\bibfnamefont {M.}~\bibnamefont
  {Cavarroc}}, \bibinfo {author} {\bibfnamefont {Y.}~\bibnamefont {Tessier}}, \
  and\ \bibinfo {author} {\bibfnamefont {L.}~\bibnamefont {Boufendi}},\
  }\href@noop {} {\bibfield  {journal} {\bibinfo  {journal} {New J. Phys.}\
  }\textbf {\bibinfo {volume} {9}},\ \bibinfo {pages} {268} (\bibinfo {year}
  {2007})}\BibitemShut {NoStop}%
\bibitem [{\citenamefont {Cou{\"e}del}\ \emph {et~al.}(2010)\citenamefont
  {Cou{\"e}del}, \citenamefont {Mikikian}, \citenamefont {Samarian},\ and\
  \citenamefont {Boufendi}}]{couedel2010self}%
  \BibitemOpen
  \bibfield  {author} {\bibinfo {author} {\bibfnamefont {L.}~\bibnamefont
  {Cou{\"e}del}}, \bibinfo {author} {\bibfnamefont {M.}~\bibnamefont
  {Mikikian}}, \bibinfo {author} {\bibfnamefont {A.~A.}\ \bibnamefont
  {Samarian}}, \ and\ \bibinfo {author} {\bibfnamefont {L.}~\bibnamefont
  {Boufendi}},\ }\href@noop {} {\bibfield  {journal} {\bibinfo  {journal}
  {Phys. Plasmas}\ }\textbf {\bibinfo {volume} {17}},\ \bibinfo {pages}
  {083705} (\bibinfo {year} {2010})}\BibitemShut {NoStop}%
\bibitem [{\citenamefont {Van~de Wetering}\ \emph {et~al.}(2015)\citenamefont
  {Van~de Wetering}, \citenamefont {Brooimans}, \citenamefont {Nijdam},
  \citenamefont {Beckers},\ and\ \citenamefont {Kroesen}}]{van2015fast}%
  \BibitemOpen
  \bibfield  {author} {\bibinfo {author} {\bibfnamefont {F.~M. J.~H.}\
  \bibnamefont {Van~de Wetering}}, \bibinfo {author} {\bibfnamefont {R.~J.~C.}\
  \bibnamefont {Brooimans}}, \bibinfo {author} {\bibfnamefont {S.}~\bibnamefont
  {Nijdam}}, \bibinfo {author} {\bibfnamefont {J.}~\bibnamefont {Beckers}}, \
  and\ \bibinfo {author} {\bibfnamefont {G.~M.~W.}\ \bibnamefont {Kroesen}},\
  }\href@noop {} {\bibfield  {journal} {\bibinfo  {journal} {J. Phys. D: Appl.
  Phys.}\ }\textbf {\bibinfo {volume} {48}},\ \bibinfo {pages} {035204}
  (\bibinfo {year} {2015})}\BibitemShut {NoStop}%
\bibitem [{\citenamefont {Wegner}\ \emph {et~al.}(2016)\citenamefont {Wegner},
  \citenamefont {Hinz}, \citenamefont {Faupel}, \citenamefont {Strunskus},
  \citenamefont {Kersten},\ and\ \citenamefont
  {Meichsner}}]{wegner2016influence}%
  \BibitemOpen
  \bibfield  {author} {\bibinfo {author} {\bibfnamefont {T.}~\bibnamefont
  {Wegner}}, \bibinfo {author} {\bibfnamefont {A.~M.}\ \bibnamefont {Hinz}},
  \bibinfo {author} {\bibfnamefont {F.}~\bibnamefont {Faupel}}, \bibinfo
  {author} {\bibfnamefont {T.}~\bibnamefont {Strunskus}}, \bibinfo {author}
  {\bibfnamefont {H.}~\bibnamefont {Kersten}}, \ and\ \bibinfo {author}
  {\bibfnamefont {J.}~\bibnamefont {Meichsner}},\ }\href@noop {} {\bibfield
  {journal} {\bibinfo  {journal} {Appl. Phys. Lett.}\ }\textbf {\bibinfo
  {volume} {108}},\ \bibinfo {pages} {063108} (\bibinfo {year}
  {2016})}\BibitemShut {NoStop}%
\bibitem [{\citenamefont {van~de Wetering}\ \emph {et~al.}(2016)\citenamefont
  {van~de Wetering}, \citenamefont {Nijdam},\ and\ \citenamefont
  {Beckers}}]{van2016conclusive}%
  \BibitemOpen
  \bibfield  {author} {\bibinfo {author} {\bibfnamefont {F.~M. J.~H.}\
  \bibnamefont {van~de Wetering}}, \bibinfo {author} {\bibfnamefont
  {S.}~\bibnamefont {Nijdam}}, \ and\ \bibinfo {author} {\bibfnamefont
  {J.}~\bibnamefont {Beckers}},\ }\href@noop {} {\bibfield  {journal} {\bibinfo
   {journal} {Appl. Phys. Lett.}\ }\textbf {\bibinfo {volume} {109}},\ \bibinfo
  {pages} {043105} (\bibinfo {year} {2016})}\BibitemShut {NoStop}%
\bibitem [{\citenamefont {Pilch}\ and\ \citenamefont
  {Greiner}(2017)}]{pilch2017diagnostics}%
  \BibitemOpen
  \bibfield  {author} {\bibinfo {author} {\bibfnamefont {I.}~\bibnamefont
  {Pilch}}\ and\ \bibinfo {author} {\bibfnamefont {F.}~\bibnamefont
  {Greiner}},\ }\href@noop {} {\bibfield  {journal} {\bibinfo  {journal} {J.
  Appl. Phys.}\ }\textbf {\bibinfo {volume} {121}},\ \bibinfo {pages} {113302}
  (\bibinfo {year} {2017})}\BibitemShut {NoStop}%
\bibitem [{\citenamefont {B\"{o}hm}\ and\ \citenamefont
  {Perrin}(1991)}]{boehm1991silanerf}%
  \BibitemOpen
  \bibfield  {author} {\bibinfo {author} {\bibfnamefont {C.}~\bibnamefont
  {B\"{o}hm}}\ and\ \bibinfo {author} {\bibfnamefont {J.}~\bibnamefont
  {Perrin}},\ }\href@noop {} {\bibfield  {journal} {\bibinfo  {journal} {J.
  Phys. D}\ }\textbf {\bibinfo {volume} {24}},\ \bibinfo {pages} {865}
  (\bibinfo {year} {1991})}\BibitemShut {NoStop}%
\bibitem [{\citenamefont {Bouchoule}\ and\ \citenamefont
  {Boufendi}(1993)}]{bouchoule1993highconcentration}%
  \BibitemOpen
  \bibfield  {author} {\bibinfo {author} {\bibfnamefont {A.}~\bibnamefont
  {Bouchoule}}\ and\ \bibinfo {author} {\bibfnamefont {L.}~\bibnamefont
  {Boufendi}},\ }\href@noop {} {\bibfield  {journal} {\bibinfo  {journal}
  {Plasma Sources Sci. Technol.}\ }\textbf {\bibinfo {volume} {2}},\ \bibinfo
  {pages} {204} (\bibinfo {year} {1993})}\BibitemShut {NoStop}%
\bibitem [{\citenamefont {Fridman}\ \emph {et~al.}(1996)\citenamefont
  {Fridman}, \citenamefont {Boufendi}, \citenamefont {Hbid}, \citenamefont
  {Potapkin},\ and\ \citenamefont {Bouchoule}}]{fridman1996dustgrowththeory}%
  \BibitemOpen
  \bibfield  {author} {\bibinfo {author} {\bibfnamefont {A.}~\bibnamefont
  {Fridman}}, \bibinfo {author} {\bibfnamefont {L.}~\bibnamefont {Boufendi}},
  \bibinfo {author} {\bibfnamefont {T.}~\bibnamefont {Hbid}}, \bibinfo {author}
  {\bibfnamefont {B.}~\bibnamefont {Potapkin}}, \ and\ \bibinfo {author}
  {\bibfnamefont {A.}~\bibnamefont {Bouchoule}},\ }\href@noop {} {\bibfield
  {journal} {\bibinfo  {journal} {J. Appl. Phys.}\ }\textbf {\bibinfo {volume}
  {79}},\ \bibinfo {pages} {1303} (\bibinfo {year} {1996})}\BibitemShut
  {NoStop}%
\bibitem [{\citenamefont {Morfill}\ and\ \citenamefont
  {Ivlev}(2009)}]{morfill2009complex}%
  \BibitemOpen
  \bibfield  {author} {\bibinfo {author} {\bibfnamefont {G.~E.}\ \bibnamefont
  {Morfill}}\ and\ \bibinfo {author} {\bibfnamefont {A.~V.}\ \bibnamefont
  {Ivlev}},\ }\href@noop {} {\bibfield  {journal} {\bibinfo  {journal} {Rev.
  Mod. Phys.}\ }\textbf {\bibinfo {volume} {81}},\ \bibinfo {pages} {1353}
  (\bibinfo {year} {2009})}\BibitemShut {NoStop}%
\bibitem [{\citenamefont {Bonitz}\ \emph {et~al.}(2010)\citenamefont {Bonitz},
  \citenamefont {Henning},\ and\ \citenamefont {Block}}]{bonitz2010complex}%
  \BibitemOpen
  \bibfield  {author} {\bibinfo {author} {\bibfnamefont {M.}~\bibnamefont
  {Bonitz}}, \bibinfo {author} {\bibfnamefont {C.}~\bibnamefont {Henning}}, \
  and\ \bibinfo {author} {\bibfnamefont {D.}~\bibnamefont {Block}},\
  }\href@noop {} {\bibfield  {journal} {\bibinfo  {journal} {Rep. Prog. Phys.}\
  }\textbf {\bibinfo {volume} {73}},\ \bibinfo {pages} {066501} (\bibinfo
  {year} {2010})}\BibitemShut {NoStop}%
\bibitem [{\citenamefont {Thomas}\ \emph {et~al.}(2008)\citenamefont {Thomas},
  \citenamefont {Morfill}, \citenamefont {Fortov}, \citenamefont {Ivlev},
  \citenamefont {Molotkov}, \citenamefont {Lipaev}, \citenamefont {Hagl},
  \citenamefont {Rothermel}, \citenamefont {Khrapak}, \citenamefont
  {Suetterlin}, \citenamefont {Rubin-Zuzic}, \citenamefont {Petrov},
  \citenamefont {Tokarev},\ and\ \citenamefont {Krikalev}}]{thomas2008complex}%
  \BibitemOpen
  \bibfield  {author} {\bibinfo {author} {\bibfnamefont {H.~M.}\ \bibnamefont
  {Thomas}}, \bibinfo {author} {\bibfnamefont {G.~E.}\ \bibnamefont {Morfill}},
  \bibinfo {author} {\bibfnamefont {V.~E.}\ \bibnamefont {Fortov}}, \bibinfo
  {author} {\bibfnamefont {A.~V.}\ \bibnamefont {Ivlev}}, \bibinfo {author}
  {\bibfnamefont {V.~I.}\ \bibnamefont {Molotkov}}, \bibinfo {author}
  {\bibfnamefont {A.~M.}\ \bibnamefont {Lipaev}}, \bibinfo {author}
  {\bibfnamefont {T.}~\bibnamefont {Hagl}}, \bibinfo {author} {\bibfnamefont
  {H.}~\bibnamefont {Rothermel}}, \bibinfo {author} {\bibfnamefont {S.~A.}\
  \bibnamefont {Khrapak}}, \bibinfo {author} {\bibfnamefont {R.~K.}\
  \bibnamefont {Suetterlin}}, \bibinfo {author} {\bibfnamefont
  {M.}~\bibnamefont {Rubin-Zuzic}}, \bibinfo {author} {\bibfnamefont {O.~F.}\
  \bibnamefont {Petrov}}, \bibinfo {author} {\bibfnamefont {V.~I.}\
  \bibnamefont {Tokarev}}, \ and\ \bibinfo {author} {\bibfnamefont {S.~K.}\
  \bibnamefont {Krikalev}},\ }\href@noop {} {\bibfield  {journal} {\bibinfo
  {journal} {New J. Phys.}\ }\textbf {\bibinfo {volume} {10}},\ \bibinfo
  {pages} {033036} (\bibinfo {year} {2008})}\BibitemShut {NoStop}%
\bibitem [{\citenamefont {Pustylnik}\ \emph {et~al.}(2016)\citenamefont
  {Pustylnik}, \citenamefont {Fink}, \citenamefont {Nosenko}, \citenamefont
  {Antonova}, \citenamefont {Hagl}, \citenamefont {Thomas}, \citenamefont
  {Zobnin}, \citenamefont {Lipaev}, \citenamefont {Usachev}, \citenamefont
  {Molotkov}, \citenamefont {Petrov}, \citenamefont {Fortov}, \citenamefont
  {Rau}, \citenamefont {Deysenroth}, \citenamefont {Albrecht}, \citenamefont
  {Kretschmer}, \citenamefont {Thoma}, \citenamefont {Morfill}, \citenamefont
  {Seurig}, \citenamefont {Stettner}, \citenamefont {Alyamovskaya},
  \citenamefont {Orr}, \citenamefont {Kufner}, \citenamefont {Lavrenko},
  \citenamefont {Padalka}, \citenamefont {Serova}, \citenamefont
  {Samokutyaev},\ and\ \citenamefont {Christoforetti}}]{pustylnik2016pk4}%
  \BibitemOpen
  \bibfield  {author} {\bibinfo {author} {\bibfnamefont {M.~Y.}\ \bibnamefont
  {Pustylnik}}, \bibinfo {author} {\bibfnamefont {M.~A.}\ \bibnamefont {Fink}},
  \bibinfo {author} {\bibfnamefont {V.}~\bibnamefont {Nosenko}}, \bibinfo
  {author} {\bibfnamefont {T.}~\bibnamefont {Antonova}}, \bibinfo {author}
  {\bibfnamefont {T.}~\bibnamefont {Hagl}}, \bibinfo {author} {\bibfnamefont
  {H.~M.}\ \bibnamefont {Thomas}}, \bibinfo {author} {\bibfnamefont {A.~V.}\
  \bibnamefont {Zobnin}}, \bibinfo {author} {\bibfnamefont {A.~M.}\
  \bibnamefont {Lipaev}}, \bibinfo {author} {\bibfnamefont {A.~D.}\
  \bibnamefont {Usachev}}, \bibinfo {author} {\bibfnamefont {V.~I.}\
  \bibnamefont {Molotkov}}, \bibinfo {author} {\bibfnamefont {O.~F.}\
  \bibnamefont {Petrov}}, \bibinfo {author} {\bibfnamefont {V.~E.}\
  \bibnamefont {Fortov}}, \bibinfo {author} {\bibfnamefont {C.}~\bibnamefont
  {Rau}}, \bibinfo {author} {\bibfnamefont {C.}~\bibnamefont {Deysenroth}},
  \bibinfo {author} {\bibfnamefont {S.}~\bibnamefont {Albrecht}}, \bibinfo
  {author} {\bibfnamefont {M.}~\bibnamefont {Kretschmer}}, \bibinfo {author}
  {\bibfnamefont {M.~H.}\ \bibnamefont {Thoma}}, \bibinfo {author}
  {\bibfnamefont {G.~E.}\ \bibnamefont {Morfill}}, \bibinfo {author}
  {\bibfnamefont {R.}~\bibnamefont {Seurig}}, \bibinfo {author} {\bibfnamefont
  {A.}~\bibnamefont {Stettner}}, \bibinfo {author} {\bibfnamefont
  {A.}~\bibnamefont {Alyamovskaya}}, \bibinfo {author} {\bibfnamefont
  {A.}~\bibnamefont {Orr}}, \bibinfo {author} {\bibfnamefont {E.}~\bibnamefont
  {Kufner}}, \bibinfo {author} {\bibfnamefont {E.~G.}\ \bibnamefont
  {Lavrenko}}, \bibinfo {author} {\bibfnamefont {G.~I.}\ \bibnamefont
  {Padalka}}, \bibinfo {author} {\bibfnamefont {E.~O.}\ \bibnamefont {Serova}},
  \bibinfo {author} {\bibfnamefont {A.~M.}\ \bibnamefont {Samokutyaev}}, \ and\
  \bibinfo {author} {\bibfnamefont {S.}~\bibnamefont {Christoforetti}},\
  }\href@noop {} {\bibfield  {journal} {\bibinfo  {journal} {Rev. Sci.
  Instrum.}\ }\textbf {\bibinfo {volume} {87}},\ \bibinfo {pages} {093505}
  (\bibinfo {year} {2016})}\BibitemShut {NoStop}%
\bibitem [{\citenamefont {Heidemann}\ \emph {et~al.}(2011)\citenamefont
  {Heidemann}, \citenamefont {Cou{\"e}del}, \citenamefont {Zhdanov},
  \citenamefont {S\"{u}tterlin}, \citenamefont {Schwabe}, \citenamefont
  {Thomas}, \citenamefont {Ivlev}, \citenamefont {Hagl}, \citenamefont
  {Morfill}, \citenamefont {Fortov}, \citenamefont {Petrov}, \citenamefont
  {Lipaev}, \citenamefont {Tokarev}, \citenamefont {Reiter},\ and\
  \citenamefont {Vinogradov}}]{heidemann2011hb}%
  \BibitemOpen
  \bibfield  {author} {\bibinfo {author} {\bibfnamefont {R.}~\bibnamefont
  {Heidemann}}, \bibinfo {author} {\bibfnamefont {L.}~\bibnamefont
  {Cou{\"e}del}}, \bibinfo {author} {\bibfnamefont {S.~K.}\ \bibnamefont
  {Zhdanov}}, \bibinfo {author} {\bibfnamefont {R.~K.}\ \bibnamefont
  {S\"{u}tterlin}}, \bibinfo {author} {\bibfnamefont {M.}~\bibnamefont
  {Schwabe}}, \bibinfo {author} {\bibfnamefont {H.~M.}\ \bibnamefont {Thomas}},
  \bibinfo {author} {\bibfnamefont {A.~V.}\ \bibnamefont {Ivlev}}, \bibinfo
  {author} {\bibfnamefont {T.}~\bibnamefont {Hagl}}, \bibinfo {author}
  {\bibfnamefont {G.~E.}\ \bibnamefont {Morfill}}, \bibinfo {author}
  {\bibfnamefont {V.~E.}\ \bibnamefont {Fortov}}, \bibinfo {author}
  {\bibfnamefont {O.~F.}\ \bibnamefont {Petrov}}, \bibinfo {author}
  {\bibfnamefont {A.~M.}\ \bibnamefont {Lipaev}}, \bibinfo {author}
  {\bibfnamefont {V.}~\bibnamefont {Tokarev}}, \bibinfo {author} {\bibfnamefont
  {T.}~\bibnamefont {Reiter}}, \ and\ \bibinfo {author} {\bibfnamefont
  {P.}~\bibnamefont {Vinogradov}},\ }\href@noop {} {\bibfield  {journal}
  {\bibinfo  {journal} {Phys. Plasmas}\ }\textbf {\bibinfo {volume} {18}},\
  \bibinfo {pages} {053701} (\bibinfo {year} {2011})}\BibitemShut {NoStop}%
\bibitem [{\citenamefont {Pustylnik}\ \emph {et~al.}(2012)\citenamefont
  {Pustylnik}, \citenamefont {Ivlev}, \citenamefont {Sadeghi}, \citenamefont
  {Heidemann}, \citenamefont {Mitic}, \citenamefont {Thomas},\ and\
  \citenamefont {Morfill}}]{pustylnik2012heterogeneous}%
  \BibitemOpen
  \bibfield  {author} {\bibinfo {author} {\bibfnamefont {M.~Y.}\ \bibnamefont
  {Pustylnik}}, \bibinfo {author} {\bibfnamefont {A.~V.}\ \bibnamefont
  {Ivlev}}, \bibinfo {author} {\bibfnamefont {N.}~\bibnamefont {Sadeghi}},
  \bibinfo {author} {\bibfnamefont {R.}~\bibnamefont {Heidemann}}, \bibinfo
  {author} {\bibfnamefont {S.}~\bibnamefont {Mitic}}, \bibinfo {author}
  {\bibfnamefont {H.~M.}\ \bibnamefont {Thomas}}, \ and\ \bibinfo {author}
  {\bibfnamefont {G.~E.}\ \bibnamefont {Morfill}},\ }\href@noop {} {\bibfield
  {journal} {\bibinfo  {journal} {Phys. Plasmas}\ }\textbf {\bibinfo {volume}
  {19}},\ \bibinfo {pages} {103701} (\bibinfo {year} {2012})}\BibitemShut
  {NoStop}%
\bibitem [{\citenamefont {H{\"u}bner}\ and\ \citenamefont
  {Melzer}(2009)}]{hubner2009dust}%
  \BibitemOpen
  \bibfield  {author} {\bibinfo {author} {\bibfnamefont {S.}~\bibnamefont
  {H{\"u}bner}}\ and\ \bibinfo {author} {\bibfnamefont {A.}~\bibnamefont
  {Melzer}},\ }\href@noop {} {\bibfield  {journal} {\bibinfo  {journal} {Phys.
  Rev. Lett.}\ }\textbf {\bibinfo {volume} {102}},\ \bibinfo {pages} {215001}
  (\bibinfo {year} {2009})}\BibitemShut {NoStop}%
\bibitem [{\citenamefont {Mitic}\ \emph {et~al.}(2009)\citenamefont {Mitic},
  \citenamefont {Pustylnik},\ and\ \citenamefont
  {Morfill}}]{mitic2009spectroscopic}%
  \BibitemOpen
  \bibfield  {author} {\bibinfo {author} {\bibfnamefont {S.}~\bibnamefont
  {Mitic}}, \bibinfo {author} {\bibfnamefont {M.~Y.}\ \bibnamefont
  {Pustylnik}}, \ and\ \bibinfo {author} {\bibfnamefont {G.~E.}\ \bibnamefont
  {Morfill}},\ }\href@noop {} {\bibfield  {journal} {\bibinfo  {journal} {New
  J. Phys.}\ }\textbf {\bibinfo {volume} {11}},\ \bibinfo {pages} {083020}
  (\bibinfo {year} {2009})}\BibitemShut {NoStop}%
\bibitem [{\citenamefont {Melzer}\ \emph {et~al.}(2011)\citenamefont {Melzer},
  \citenamefont {H{\"u}bner}, \citenamefont {Lewerentz}, \citenamefont
  {Matyash}, \citenamefont {Schneider},\ and\ \citenamefont
  {Ikkurthi}}]{melzer2011phase}%
  \BibitemOpen
  \bibfield  {author} {\bibinfo {author} {\bibfnamefont {A.}~\bibnamefont
  {Melzer}}, \bibinfo {author} {\bibfnamefont {S.}~\bibnamefont {H{\"u}bner}},
  \bibinfo {author} {\bibfnamefont {L.}~\bibnamefont {Lewerentz}}, \bibinfo
  {author} {\bibfnamefont {K.}~\bibnamefont {Matyash}}, \bibinfo {author}
  {\bibfnamefont {R.}~\bibnamefont {Schneider}}, \ and\ \bibinfo {author}
  {\bibfnamefont {R.}~\bibnamefont {Ikkurthi}},\ }\href@noop {} {\bibfield
  {journal} {\bibinfo  {journal} {Phys. Rev. E}\ }\textbf {\bibinfo {volume}
  {83}},\ \bibinfo {pages} {036411} (\bibinfo {year} {2011})}\BibitemShut
  {NoStop}%
\bibitem [{\citenamefont {Killer}\ \emph {et~al.}(2013)\citenamefont {Killer},
  \citenamefont {Bandelow}, \citenamefont {Matyash}, \citenamefont
  {Schneider},\ and\ \citenamefont {Melzer}}]{killer2013observation}%
  \BibitemOpen
  \bibfield  {author} {\bibinfo {author} {\bibfnamefont {C.}~\bibnamefont
  {Killer}}, \bibinfo {author} {\bibfnamefont {G.}~\bibnamefont {Bandelow}},
  \bibinfo {author} {\bibfnamefont {K.}~\bibnamefont {Matyash}}, \bibinfo
  {author} {\bibfnamefont {R.}~\bibnamefont {Schneider}}, \ and\ \bibinfo
  {author} {\bibfnamefont {A.}~\bibnamefont {Melzer}},\ }\href@noop {}
  {\bibfield  {journal} {\bibinfo  {journal} {Phys. Plasmas}\ }\textbf
  {\bibinfo {volume} {20}},\ \bibinfo {pages} {083704} (\bibinfo {year}
  {2013})}\BibitemShut {NoStop}%
\bibitem [{\citenamefont {Schulze}\ \emph {et~al.}(2011)\citenamefont
  {Schulze}, \citenamefont {Derzsi}, \citenamefont {Dittmann}, \citenamefont
  {Hemke}, \citenamefont {Meichsner},\ and\ \citenamefont
  {Donk{\'o}}}]{schulze2011ionization}%
  \BibitemOpen
  \bibfield  {author} {\bibinfo {author} {\bibfnamefont {J.}~\bibnamefont
  {Schulze}}, \bibinfo {author} {\bibfnamefont {A.}~\bibnamefont {Derzsi}},
  \bibinfo {author} {\bibfnamefont {K.}~\bibnamefont {Dittmann}}, \bibinfo
  {author} {\bibfnamefont {T.}~\bibnamefont {Hemke}}, \bibinfo {author}
  {\bibfnamefont {J.}~\bibnamefont {Meichsner}}, \ and\ \bibinfo {author}
  {\bibfnamefont {Z.}~\bibnamefont {Donk{\'o}}},\ }\href@noop {} {\bibfield
  {journal} {\bibinfo  {journal} {Phys. Rev. Lett.}\ }\textbf {\bibinfo
  {volume} {107}},\ \bibinfo {pages} {275001} (\bibinfo {year}
  {2011})}\BibitemShut {NoStop}%
\bibitem [{\citenamefont {Sch{\"u}ngel}\ \emph {et~al.}(2013)\citenamefont
  {Sch{\"u}ngel}, \citenamefont {Mohr}, \citenamefont {Iwashita}, \citenamefont
  {Schulze},\ and\ \citenamefont {Czarnetzki}}]{schungel2013effect}%
  \BibitemOpen
  \bibfield  {author} {\bibinfo {author} {\bibfnamefont {E.}~\bibnamefont
  {Sch{\"u}ngel}}, \bibinfo {author} {\bibfnamefont {S.}~\bibnamefont {Mohr}},
  \bibinfo {author} {\bibfnamefont {S.}~\bibnamefont {Iwashita}}, \bibinfo
  {author} {\bibfnamefont {J.}~\bibnamefont {Schulze}}, \ and\ \bibinfo
  {author} {\bibfnamefont {U.}~\bibnamefont {Czarnetzki}},\ }\href@noop {}
  {\bibfield  {journal} {\bibinfo  {journal} {J. Phys. D: Appl. Phys.}\
  }\textbf {\bibinfo {volume} {46}},\ \bibinfo {pages} {175205} (\bibinfo
  {year} {2013})}\BibitemShut {NoStop}%
\bibitem [{\citenamefont {Hemke}\ \emph {et~al.}(2012)\citenamefont {Hemke},
  \citenamefont {Eremin}, \citenamefont {Mussenbrock}, \citenamefont {Derzsi},
  \citenamefont {Donk{\'o}}, \citenamefont {Dittmann}, \citenamefont
  {Meichsner},\ and\ \citenamefont {Schulze}}]{hemke2012ionization}%
  \BibitemOpen
  \bibfield  {author} {\bibinfo {author} {\bibfnamefont {T.}~\bibnamefont
  {Hemke}}, \bibinfo {author} {\bibfnamefont {D.}~\bibnamefont {Eremin}},
  \bibinfo {author} {\bibfnamefont {T.}~\bibnamefont {Mussenbrock}}, \bibinfo
  {author} {\bibfnamefont {A.}~\bibnamefont {Derzsi}}, \bibinfo {author}
  {\bibfnamefont {Z.}~\bibnamefont {Donk{\'o}}}, \bibinfo {author}
  {\bibfnamefont {K.}~\bibnamefont {Dittmann}}, \bibinfo {author}
  {\bibfnamefont {J.}~\bibnamefont {Meichsner}}, \ and\ \bibinfo {author}
  {\bibfnamefont {J.}~\bibnamefont {Schulze}},\ }\href@noop {} {\bibfield
  {journal} {\bibinfo  {journal} {Plasma Sources Sci. Technol.}\ }\textbf
  {\bibinfo {volume} {22}},\ \bibinfo {pages} {015012} (\bibinfo {year}
  {2012})}\BibitemShut {NoStop}%
\bibitem [{\citenamefont {Rothermel}\ \emph {et~al.}(2002)\citenamefont
  {Rothermel}, \citenamefont {Hagl}, \citenamefont {Morfill}, \citenamefont
  {Thoma},\ and\ \citenamefont {Thomas}}]{rothermel2002thermophoresis}%
  \BibitemOpen
  \bibfield  {author} {\bibinfo {author} {\bibfnamefont {H.}~\bibnamefont
  {Rothermel}}, \bibinfo {author} {\bibfnamefont {T.}~\bibnamefont {Hagl}},
  \bibinfo {author} {\bibfnamefont {G.~E.}\ \bibnamefont {Morfill}}, \bibinfo
  {author} {\bibfnamefont {M.~H.}\ \bibnamefont {Thoma}}, \ and\ \bibinfo
  {author} {\bibfnamefont {H.~M.}\ \bibnamefont {Thomas}},\ }\href@noop {}
  {\bibfield  {journal} {\bibinfo  {journal} {Phys. Rev. Lett.}\ }\textbf
  {\bibinfo {volume} {89}} (\bibinfo {year} {2002})}\BibitemShut {NoStop}%
\bibitem [{\citenamefont {Schwabe}\ \emph {et~al.}(2007)\citenamefont
  {Schwabe}, \citenamefont {Rubin-Zuzic}, \citenamefont {Zhdanov},
  \citenamefont {Thomas},\ and\ \citenamefont {Morfill}}]{schwabe2007highly}%
  \BibitemOpen
  \bibfield  {author} {\bibinfo {author} {\bibfnamefont {M.}~\bibnamefont
  {Schwabe}}, \bibinfo {author} {\bibfnamefont {M.}~\bibnamefont
  {Rubin-Zuzic}}, \bibinfo {author} {\bibfnamefont {S.}~\bibnamefont
  {Zhdanov}}, \bibinfo {author} {\bibfnamefont {H.~M.}\ \bibnamefont {Thomas}},
  \ and\ \bibinfo {author} {\bibfnamefont {G.~E.}\ \bibnamefont {Morfill}},\
  }\href@noop {} {\bibfield  {journal} {\bibinfo  {journal} {Phys. Rev. Lett.}\
  }\textbf {\bibinfo {volume} {99}},\ \bibinfo {pages} {095002} (\bibinfo
  {year} {2007})}\BibitemShut {NoStop}%
\bibitem [{\citenamefont {Wiese}\ \emph {et~al.}(1989)\citenamefont {Wiese},
  \citenamefont {Brault}, \citenamefont {Danzmann}, \citenamefont {Helbig},\
  and\ \citenamefont {Kock}}]{wiese1989unified}%
  \BibitemOpen
  \bibfield  {author} {\bibinfo {author} {\bibfnamefont {W.~L.}\ \bibnamefont
  {Wiese}}, \bibinfo {author} {\bibfnamefont {J.~W.}\ \bibnamefont {Brault}},
  \bibinfo {author} {\bibfnamefont {K.}~\bibnamefont {Danzmann}}, \bibinfo
  {author} {\bibfnamefont {V.}~\bibnamefont {Helbig}}, \ and\ \bibinfo {author}
  {\bibfnamefont {M.}~\bibnamefont {Kock}},\ }\href@noop {} {\bibfield
  {journal} {\bibinfo  {journal} {Phys. Rev. A}\ }\textbf {\bibinfo {volume}
  {39}},\ \bibinfo {pages} {2461} (\bibinfo {year} {1989})}\BibitemShut
  {NoStop}%
\bibitem [{\citenamefont {Vahedi}\ and\ \citenamefont
  {Surendra}(1995)}]{vahedi1995monte}%
  \BibitemOpen
  \bibfield  {author} {\bibinfo {author} {\bibfnamefont {V.}~\bibnamefont
  {Vahedi}}\ and\ \bibinfo {author} {\bibfnamefont {M.}~\bibnamefont
  {Surendra}},\ }\href@noop {} {\bibfield  {journal} {\bibinfo  {journal}
  {Comput. Phys. Commun.}\ }\textbf {\bibinfo {volume} {87}},\ \bibinfo {pages}
  {179} (\bibinfo {year} {1995})}\BibitemShut {NoStop}%
\bibitem [{\citenamefont {Donk{\'o}}(2011)}]{donko2011particle}%
  \BibitemOpen
  \bibfield  {author} {\bibinfo {author} {\bibfnamefont {Z.}~\bibnamefont
  {Donk{\'o}}},\ }\href@noop {} {\bibfield  {journal} {\bibinfo  {journal}
  {Plasma Sources Sci. Technol.}\ }\textbf {\bibinfo {volume} {20}},\ \bibinfo
  {pages} {024001} (\bibinfo {year} {2011})}\BibitemShut {NoStop}%
\bibitem [{\citenamefont {Semenov}(2017)}]{semenov2016moment}%
  \BibitemOpen
  \bibfield  {author} {\bibinfo {author} {\bibfnamefont {I.~L.}\ \bibnamefont
  {Semenov}},\ }\href {\doibase 10.1103/PhysRevE.95.043208} {\bibfield
  {journal} {\bibinfo  {journal} {Phys. Rev. E}\ }\textbf {\bibinfo {volume}
  {95}},\ \bibinfo {pages} {043208} (\bibinfo {year} {2017})}\BibitemShut
  {NoStop}%
\bibitem [{\citenamefont {Gozadinos}\ \emph {et~al.}(2003)\citenamefont
  {Gozadinos}, \citenamefont {Ivlev},\ and\ \citenamefont
  {Boeuf}}]{gozadinos2003fluid}%
  \BibitemOpen
  \bibfield  {author} {\bibinfo {author} {\bibfnamefont {G.}~\bibnamefont
  {Gozadinos}}, \bibinfo {author} {\bibfnamefont {A.~V.}\ \bibnamefont
  {Ivlev}}, \ and\ \bibinfo {author} {\bibfnamefont {J.~P.}\ \bibnamefont
  {Boeuf}},\ }\href@noop {} {\bibfield  {journal} {\bibinfo  {journal} {New J.
  Phys.}\ }\textbf {\bibinfo {volume} {5}},\ \bibinfo {pages} {32} (\bibinfo
  {year} {2003})}\BibitemShut {NoStop}%
\bibitem [{\citenamefont {Land}\ and\ \citenamefont
  {Goedheer}(2006)}]{land2006effect}%
  \BibitemOpen
  \bibfield  {author} {\bibinfo {author} {\bibfnamefont {V.}~\bibnamefont
  {Land}}\ and\ \bibinfo {author} {\bibfnamefont {W.~J.}\ \bibnamefont
  {Goedheer}},\ }\href@noop {} {\bibfield  {journal} {\bibinfo  {journal} {New
  J. Phys.}\ }\textbf {\bibinfo {volume} {8}},\ \bibinfo {pages} {8} (\bibinfo
  {year} {2006})}\BibitemShut {NoStop}%
\bibitem [{\citenamefont {Land}\ and\ \citenamefont
  {Goedheer}(2007)}]{land2007plasma}%
  \BibitemOpen
  \bibfield  {author} {\bibinfo {author} {\bibfnamefont {V.}~\bibnamefont
  {Land}}\ and\ \bibinfo {author} {\bibfnamefont {W.~J.}\ \bibnamefont
  {Goedheer}},\ }\href@noop {} {\bibfield  {journal} {\bibinfo  {journal} {New
  J. Phys.}\ }\textbf {\bibinfo {volume} {9}},\ \bibinfo {pages} {246}
  (\bibinfo {year} {2007})}\BibitemShut {NoStop}%
\bibitem [{\citenamefont {Goedheer}\ \emph {et~al.}(2009)\citenamefont
  {Goedheer}, \citenamefont {Land},\ and\ \citenamefont
  {Venema}}]{goedheer2009hydrodynamic}%
  \BibitemOpen
  \bibfield  {author} {\bibinfo {author} {\bibfnamefont {W.~J.}\ \bibnamefont
  {Goedheer}}, \bibinfo {author} {\bibfnamefont {V.}~\bibnamefont {Land}}, \
  and\ \bibinfo {author} {\bibfnamefont {J.}~\bibnamefont {Venema}},\
  }\href@noop {} {\bibfield  {journal} {\bibinfo  {journal} {J. Phys. D: Appl.
  Phys.}\ }\textbf {\bibinfo {volume} {42}},\ \bibinfo {pages} {194015}
  (\bibinfo {year} {2009})}\BibitemShut {NoStop}%
\bibitem [{\citenamefont {Land}\ \emph {et~al.}(2010)\citenamefont {Land},
  \citenamefont {Matthews}, \citenamefont {Hyde},\ and\ \citenamefont
  {Bolser}}]{land2010fluid}%
  \BibitemOpen
  \bibfield  {author} {\bibinfo {author} {\bibfnamefont {V.}~\bibnamefont
  {Land}}, \bibinfo {author} {\bibfnamefont {L.~S.}\ \bibnamefont {Matthews}},
  \bibinfo {author} {\bibfnamefont {T.~W.}\ \bibnamefont {Hyde}}, \ and\
  \bibinfo {author} {\bibfnamefont {D.}~\bibnamefont {Bolser}},\ }\href@noop {}
  {\bibfield  {journal} {\bibinfo  {journal} {Phys. Rev. E}\ }\textbf {\bibinfo
  {volume} {81}},\ \bibinfo {pages} {056402} (\bibinfo {year}
  {2010})}\BibitemShut {NoStop}%
\bibitem [{\citenamefont {Fortov}\ \emph {et~al.}(2005)\citenamefont {Fortov},
  \citenamefont {Ivlev}, \citenamefont {Khrapak}, \citenamefont {Khrapak},\
  and\ \citenamefont {Morfill}}]{fortov2005complex}%
  \BibitemOpen
  \bibfield  {author} {\bibinfo {author} {\bibfnamefont {V.~E.}\ \bibnamefont
  {Fortov}}, \bibinfo {author} {\bibfnamefont {A.~V.}\ \bibnamefont {Ivlev}},
  \bibinfo {author} {\bibfnamefont {S.~A.}\ \bibnamefont {Khrapak}}, \bibinfo
  {author} {\bibfnamefont {A.~G.}\ \bibnamefont {Khrapak}}, \ and\ \bibinfo
  {author} {\bibfnamefont {G.~E.}\ \bibnamefont {Morfill}},\ }\href@noop {}
  {\bibfield  {journal} {\bibinfo  {journal} {Phys. Rep.}\ }\textbf {\bibinfo
  {volume} {421}},\ \bibinfo {pages} {1} (\bibinfo {year} {2005})}\BibitemShut
  {NoStop}%
\bibitem [{\citenamefont {Landau}\ and\ \citenamefont
  {Lifshitz}(1965)}]{landau1965course}%
  \BibitemOpen
  \bibfield  {author} {\bibinfo {author} {\bibfnamefont {L.~D.}\ \bibnamefont
  {Landau}}\ and\ \bibinfo {author} {\bibfnamefont {E.~M.}\ \bibnamefont
  {Lifshitz}},\ }\href@noop {} {\emph {\bibinfo {title} {Course of Theoretical
  Physics. Volume 3: Quantum Mechanics}}}\ (\bibinfo  {publisher} {Pergamon},\
  \bibinfo {year} {1965})\BibitemShut {NoStop}%
\bibitem [{\citenamefont {McEachran}\ and\ \citenamefont
  {Stauffer}(1983)}]{mceachran1983elastic}%
  \BibitemOpen
  \bibfield  {author} {\bibinfo {author} {\bibfnamefont {R.~P.}\ \bibnamefont
  {McEachran}}\ and\ \bibinfo {author} {\bibfnamefont {A.~D.}\ \bibnamefont
  {Stauffer}},\ }\href@noop {} {\bibfield  {journal} {\bibinfo  {journal} {J.
  Phys. B}\ }\textbf {\bibinfo {volume} {16}},\ \bibinfo {pages} {4023}
  (\bibinfo {year} {1983})}\BibitemShut {NoStop}%
\bibitem [{\citenamefont {Khrapak}\ and\ \citenamefont
  {Morfill}(2008)}]{khrapak2008interpolation}%
  \BibitemOpen
  \bibfield  {author} {\bibinfo {author} {\bibfnamefont {S.~A.}\ \bibnamefont
  {Khrapak}}\ and\ \bibinfo {author} {\bibfnamefont {G.~E.}\ \bibnamefont
  {Morfill}},\ }\href@noop {} {\bibfield  {journal} {\bibinfo  {journal} {Phys.
  Plasmas}\ }\textbf {\bibinfo {volume} {15}},\ \bibinfo {pages} {114503}
  (\bibinfo {year} {2008})}\BibitemShut {NoStop}%
\bibitem [{\citenamefont {Khrapak}(2014)}]{khrapak2014accurate}%
  \BibitemOpen
  \bibfield  {author} {\bibinfo {author} {\bibfnamefont {S.~A.}\ \bibnamefont
  {Khrapak}},\ }\href@noop {} {\bibfield  {journal} {\bibinfo  {journal} {Phys.
  Plasmas}\ }\textbf {\bibinfo {volume} {21}},\ \bibinfo {pages} {044506}
  (\bibinfo {year} {2014})}\BibitemShut {NoStop}%
\bibitem [{\citenamefont {Ratynskaia}\ \emph {et~al.}(2006)\citenamefont
  {Ratynskaia}, \citenamefont {De~Angelis}, \citenamefont {Khrapak},
  \citenamefont {Klumov},\ and\ \citenamefont
  {Morfill}}]{ratynskaia2006electrostatic}%
  \BibitemOpen
  \bibfield  {author} {\bibinfo {author} {\bibfnamefont {S.}~\bibnamefont
  {Ratynskaia}}, \bibinfo {author} {\bibfnamefont {U.}~\bibnamefont
  {De~Angelis}}, \bibinfo {author} {\bibfnamefont {S.}~\bibnamefont {Khrapak}},
  \bibinfo {author} {\bibfnamefont {B.}~\bibnamefont {Klumov}}, \ and\ \bibinfo
  {author} {\bibfnamefont {G.~E.}\ \bibnamefont {Morfill}},\ }\href@noop {}
  {\bibfield  {journal} {\bibinfo  {journal} {Phys. Plasmas}\ }\textbf
  {\bibinfo {volume} {13}},\ \bibinfo {pages} {104508} (\bibinfo {year}
  {2006})}\BibitemShut {NoStop}%
\bibitem [{\citenamefont {Semenov}\ \emph {et~al.}(2015)\citenamefont
  {Semenov}, \citenamefont {Khrapak},\ and\ \citenamefont
  {Thomas}}]{semenov2015approximate}%
  \BibitemOpen
  \bibfield  {author} {\bibinfo {author} {\bibfnamefont {I.~L.}\ \bibnamefont
  {Semenov}}, \bibinfo {author} {\bibfnamefont {S.~A.}\ \bibnamefont
  {Khrapak}}, \ and\ \bibinfo {author} {\bibfnamefont {H.~M.}\ \bibnamefont
  {Thomas}},\ }\href@noop {} {\bibfield  {journal} {\bibinfo  {journal} {Phys.
  Plasmas}\ }\textbf {\bibinfo {volume} {22}},\ \bibinfo {pages} {053704}
  (\bibinfo {year} {2015})}\BibitemShut {NoStop}%
\bibitem [{\citenamefont {Semenov}\ \emph {et~al.}(2017)\citenamefont
  {Semenov}, \citenamefont {Khrapak},\ and\ \citenamefont
  {Thomas}}]{semenov2017momentum}%
  \BibitemOpen
  \bibfield  {author} {\bibinfo {author} {\bibfnamefont {I.~L.}\ \bibnamefont
  {Semenov}}, \bibinfo {author} {\bibfnamefont {S.~A.}\ \bibnamefont
  {Khrapak}}, \ and\ \bibinfo {author} {\bibfnamefont {H.~M.}\ \bibnamefont
  {Thomas}},\ }\href@noop {} {\bibfield  {journal} {\bibinfo  {journal} {Phys.
  Plasmas}\ }\textbf {\bibinfo {volume} {24}},\ \bibinfo {pages} {033710}
  (\bibinfo {year} {2017})}\BibitemShut {NoStop}%
\bibitem [{\citenamefont {Lampe}\ \emph {et~al.}(2012)\citenamefont {Lampe},
  \citenamefont {R{\"o}cker}, \citenamefont {Joyce}, \citenamefont {Zhdanov},
  \citenamefont {Ivlev},\ and\ \citenamefont {Morfill}}]{lampe2012ion}%
  \BibitemOpen
  \bibfield  {author} {\bibinfo {author} {\bibfnamefont {M.}~\bibnamefont
  {Lampe}}, \bibinfo {author} {\bibfnamefont {T.~B.}\ \bibnamefont
  {R{\"o}cker}}, \bibinfo {author} {\bibfnamefont {G.}~\bibnamefont {Joyce}},
  \bibinfo {author} {\bibfnamefont {S.~K.}\ \bibnamefont {Zhdanov}}, \bibinfo
  {author} {\bibfnamefont {A.~V.}\ \bibnamefont {Ivlev}}, \ and\ \bibinfo
  {author} {\bibfnamefont {G.~E.}\ \bibnamefont {Morfill}},\ }\href@noop {}
  {\bibfield  {journal} {\bibinfo  {journal} {Phys. Plasmas}\ }\textbf
  {\bibinfo {volume} {19}},\ \bibinfo {pages} {113703} (\bibinfo {year}
  {2012})}\BibitemShut {NoStop}%
\bibitem [{\citenamefont {Khrapak}\ \emph
  {et~al.}(2014{\natexlab{a}})\citenamefont {Khrapak}, \citenamefont {Khrapak},
  \citenamefont {Ivlev},\ and\ \citenamefont {Thomas}}]{khrapak2014ion}%
  \BibitemOpen
  \bibfield  {author} {\bibinfo {author} {\bibfnamefont {S.~A.}\ \bibnamefont
  {Khrapak}}, \bibinfo {author} {\bibfnamefont {A.~G.}\ \bibnamefont
  {Khrapak}}, \bibinfo {author} {\bibfnamefont {A.~V.}\ \bibnamefont {Ivlev}},
  \ and\ \bibinfo {author} {\bibfnamefont {H.~M.}\ \bibnamefont {Thomas}},\
  }\href@noop {} {\bibfield  {journal} {\bibinfo  {journal} {Phys. Plasmas}\
  }\textbf {\bibinfo {volume} {21}},\ \bibinfo {pages} {123705} (\bibinfo
  {year} {2014}{\natexlab{a}})}\BibitemShut {NoStop}%
\bibitem [{\citenamefont {Khrapak}\ \emph
  {et~al.}(2014{\natexlab{b}})\citenamefont {Khrapak}, \citenamefont {Khrapak},
  \citenamefont {Ivlev},\ and\ \citenamefont {Morfill}}]{khrapak2014simple}%
  \BibitemOpen
  \bibfield  {author} {\bibinfo {author} {\bibfnamefont {S.~A.}\ \bibnamefont
  {Khrapak}}, \bibinfo {author} {\bibfnamefont {A.~G.}\ \bibnamefont
  {Khrapak}}, \bibinfo {author} {\bibfnamefont {A.~V.}\ \bibnamefont {Ivlev}},
  \ and\ \bibinfo {author} {\bibfnamefont {G.~E.}\ \bibnamefont {Morfill}},\
  }\href@noop {} {\bibfield  {journal} {\bibinfo  {journal} {Phys. Rev. E}\
  }\textbf {\bibinfo {volume} {89}},\ \bibinfo {pages} {023102} (\bibinfo
  {year} {2014}{\natexlab{b}})}\BibitemShut {NoStop}%
\bibitem [{\citenamefont {Khrapak}\ and\ \citenamefont
  {Thomas}(2015)}]{khrapak2015practical}%
  \BibitemOpen
  \bibfield  {author} {\bibinfo {author} {\bibfnamefont {S.~A.}\ \bibnamefont
  {Khrapak}}\ and\ \bibinfo {author} {\bibfnamefont {H.~M.}\ \bibnamefont
  {Thomas}},\ }\href@noop {} {\bibfield  {journal} {\bibinfo  {journal} {Phys.
  Rev. E}\ }\textbf {\bibinfo {volume} {91}},\ \bibinfo {pages} {023108}
  (\bibinfo {year} {2015})}\BibitemShut {NoStop}%
\bibitem [{\citenamefont {LeVeque}(2002)}]{leveque2002finite}%
  \BibitemOpen
  \bibfield  {author} {\bibinfo {author} {\bibfnamefont {R.~J.}\ \bibnamefont
  {LeVeque}},\ }\href@noop {} {\emph {\bibinfo {title} {Finite Volume Methods
  for Hyperbolic Problems}}}\ (\bibinfo  {publisher} {Cambridge University
  Press},\ \bibinfo {year} {2002})\BibitemShut {NoStop}%
\bibitem [{\citenamefont {Lipaev}\ \emph {et~al.}(2007)\citenamefont {Lipaev},
  \citenamefont {Khrapak}, \citenamefont {Molotkov}, \citenamefont {Morfill},
  \citenamefont {Fortov}, \citenamefont {Ivlev}, \citenamefont {Thomas},
  \citenamefont {Khrapak}, \citenamefont {Naumkin}, \citenamefont {Ivanov},
  \citenamefont {Tretschev},\ and\ \citenamefont {Padalka}}]{lipaev2007void}%
  \BibitemOpen
  \bibfield  {author} {\bibinfo {author} {\bibfnamefont {A.~M.}\ \bibnamefont
  {Lipaev}}, \bibinfo {author} {\bibfnamefont {S.~A.}\ \bibnamefont {Khrapak}},
  \bibinfo {author} {\bibfnamefont {V.~I.}\ \bibnamefont {Molotkov}}, \bibinfo
  {author} {\bibfnamefont {G.~E.}\ \bibnamefont {Morfill}}, \bibinfo {author}
  {\bibfnamefont {V.~E.}\ \bibnamefont {Fortov}}, \bibinfo {author}
  {\bibfnamefont {A.~V.}\ \bibnamefont {Ivlev}}, \bibinfo {author}
  {\bibfnamefont {H.~M.}\ \bibnamefont {Thomas}}, \bibinfo {author}
  {\bibfnamefont {A.~G.}\ \bibnamefont {Khrapak}}, \bibinfo {author}
  {\bibfnamefont {V.~N.}\ \bibnamefont {Naumkin}}, \bibinfo {author}
  {\bibfnamefont {A.~I.}\ \bibnamefont {Ivanov}}, \bibinfo {author}
  {\bibfnamefont {S.~E.}\ \bibnamefont {Tretschev}}, \ and\ \bibinfo {author}
  {\bibfnamefont {G.~I.}\ \bibnamefont {Padalka}},\ }\href@noop {} {\bibfield
  {journal} {\bibinfo  {journal} {Phys. Rev. Lett.}\ }\textbf {\bibinfo
  {volume} {98}},\ \bibinfo {pages} {265006} (\bibinfo {year}
  {2007})}\BibitemShut {NoStop}%
\bibitem [{LXC()}]{LXCat}%
  \BibitemOpen
  \href@noop {} {}\bibinfo {howpublished}
  {\url{www.lxcat.laplace.univ-tlse.fr}}\BibitemShut {NoStop}%
\bibitem [{\citenamefont {Sukhinin}\ \emph
  {et~al.}(2013{\natexlab{a}})\citenamefont {Sukhinin}, \citenamefont
  {Fedoseev}, \citenamefont {Salnikov}, \citenamefont {Antipov}, \citenamefont
  {Petrov},\ and\ \citenamefont {Fortov}}]{sukhinin2013influence}%
  \BibitemOpen
  \bibfield  {author} {\bibinfo {author} {\bibfnamefont {G.~I.}\ \bibnamefont
  {Sukhinin}}, \bibinfo {author} {\bibfnamefont {A.~V.}\ \bibnamefont
  {Fedoseev}}, \bibinfo {author} {\bibfnamefont {M.~V.}\ \bibnamefont
  {Salnikov}}, \bibinfo {author} {\bibfnamefont {S.~N.}\ \bibnamefont
  {Antipov}}, \bibinfo {author} {\bibfnamefont {O.~F.}\ \bibnamefont {Petrov}},
  \ and\ \bibinfo {author} {\bibfnamefont {V.~E.}\ \bibnamefont {Fortov}},\
  }\href@noop {} {\bibfield  {journal} {\bibinfo  {journal} {EPL}\ }\textbf
  {\bibinfo {volume} {103}},\ \bibinfo {pages} {35001} (\bibinfo {year}
  {2013}{\natexlab{a}})}\BibitemShut {NoStop}%
\bibitem [{\citenamefont {Sukhinin}\ \emph
  {et~al.}(2013{\natexlab{b}})\citenamefont {Sukhinin}, \citenamefont
  {Fedoseev}, \citenamefont {Antipov}, \citenamefont {Petrov},\ and\
  \citenamefont {Fortov}}]{sukhinin2013dust}%
  \BibitemOpen
  \bibfield  {author} {\bibinfo {author} {\bibfnamefont {G.~I.}\ \bibnamefont
  {Sukhinin}}, \bibinfo {author} {\bibfnamefont {A.~V.}\ \bibnamefont
  {Fedoseev}}, \bibinfo {author} {\bibfnamefont {S.~N.}\ \bibnamefont
  {Antipov}}, \bibinfo {author} {\bibfnamefont {O.~F.}\ \bibnamefont {Petrov}},
  \ and\ \bibinfo {author} {\bibfnamefont {V.~E.}\ \bibnamefont {Fortov}},\
  }\href@noop {} {\bibfield  {journal} {\bibinfo  {journal} {Phys. Rev. E}\
  }\textbf {\bibinfo {volume} {87}},\ \bibinfo {pages} {013101} (\bibinfo
  {year} {2013}{\natexlab{b}})}\BibitemShut {NoStop}%
\bibitem [{\citenamefont {Ishiguro}\ \emph {et~al.}(1985)\citenamefont
  {Ishiguro}, \citenamefont {Kamimura},\ and\ \citenamefont
  {Sato}}]{ishiguro1985dblayer}%
  \BibitemOpen
  \bibfield  {author} {\bibinfo {author} {\bibfnamefont {S.}~\bibnamefont
  {Ishiguro}}, \bibinfo {author} {\bibfnamefont {T.}~\bibnamefont {Kamimura}},
  \ and\ \bibinfo {author} {\bibfnamefont {T.}~\bibnamefont {Sato}},\
  }\href@noop {} {\bibfield  {journal} {\bibinfo  {journal} {Phys. Fluids}\
  }\textbf {\bibinfo {volume} {28}},\ \bibinfo {pages} {2100} (\bibinfo {year}
  {1985})}\BibitemShut {NoStop}%
\bibitem [{\citenamefont {Raizer}\ \emph {et~al.}(1995)\citenamefont {Raizer},
  \citenamefont {Shneider},\ and\ \citenamefont
  {Yatsenko}}]{raizer1995rfdischarge}%
  \BibitemOpen
  \bibfield  {author} {\bibinfo {author} {\bibfnamefont {Y.}~\bibnamefont
  {Raizer}}, \bibinfo {author} {\bibfnamefont {M.}~\bibnamefont {Shneider}}, \
  and\ \bibinfo {author} {\bibfnamefont {N.}~\bibnamefont {Yatsenko}},\
  }\href@noop {} {\emph {\bibinfo {title} {Radio-Frequency Capacitive
  Discharges}}}\ (\bibinfo  {publisher} {CRC Press},\ \bibinfo {year}
  {1995})\BibitemShut {NoStop}%
\bibitem [{\citenamefont {Samsonov}\ and\ \citenamefont
  {Goree}(1999)}]{samsonov1999instabilities}%
  \BibitemOpen
  \bibfield  {author} {\bibinfo {author} {\bibfnamefont {D.}~\bibnamefont
  {Samsonov}}\ and\ \bibinfo {author} {\bibfnamefont {J.}~\bibnamefont
  {Goree}},\ }\href@noop {} {\bibfield  {journal} {\bibinfo  {journal} {Phys.
  Rev. E}\ }\textbf {\bibinfo {volume} {59}},\ \bibinfo {pages} {1047}
  (\bibinfo {year} {1999})}\BibitemShut {NoStop}%
\bibitem [{\citenamefont {Berndt}\ \emph {et~al.}(2005)\citenamefont {Berndt},
  \citenamefont {Kova{\v{c}}evi{\'c}}, \citenamefont {Selenin}, \citenamefont
  {Stefanovi{\'c}},\ and\ \citenamefont {Winter}}]{berndt2005anomalous}%
  \BibitemOpen
  \bibfield  {author} {\bibinfo {author} {\bibfnamefont {J.}~\bibnamefont
  {Berndt}}, \bibinfo {author} {\bibfnamefont {E.}~\bibnamefont
  {Kova{\v{c}}evi{\'c}}}, \bibinfo {author} {\bibfnamefont {V.}~\bibnamefont
  {Selenin}}, \bibinfo {author} {\bibfnamefont {I.}~\bibnamefont
  {Stefanovi{\'c}}}, \ and\ \bibinfo {author} {\bibfnamefont {J.}~\bibnamefont
  {Winter}},\ }\href@noop {} {\bibfield  {journal} {\bibinfo  {journal} {Plasma
  Sources Sci. Technol.}\ }\textbf {\bibinfo {volume} {15}},\ \bibinfo {pages}
  {18} (\bibinfo {year} {2005})}\BibitemShut {NoStop}%
\end{thebibliography}
\providecommand{\noopsort}[1]{}\providecommand{\singleletter}[1]{#1}%

\end{document}